\documentclass[12pt]{article}
\usepackage{amssymb}
\newtheorem{proposition}{Proposition}[section]
\newtheorem{remark}[proposition]{Remark}

\newtheorem{theorem}[proposition]{Theorem}
\newtheorem{lemma}[proposition]{Lemma}
\newtheorem{corollary}[proposition]{Corollary}

\newtheorem{axiom[proposition]}{Axiome}
\newcommand{\Section}[1]{\setcounter{equation}{0}\section{#1}}

\def\stackunder#1#2{\mathrel{\mathop{#2}\limits_{#1}}}%
\newcommand{\Tr}{{\rm Tr}}
\input{tcilatex}
\begin{document}

\font\fifteen=cmbx10 at 15pt \font\twelve=cmbx10 at 12pt

\begin{titlepage}

\begin{center}

\renewcommand{\thefootnote}{\fnsymbol{footnote}}

\vspace{6 cm}

{\fifteen The equilibrium states for a model with two kinds of
Bose condensation}

\vspace{4 cm}

\setcounter{footnote}{0}
\renewcommand{\thefootnote}{\arabic{footnote}}

{\bf J.-B. BRU\footnote{Department of Mathematics, UC Davis, -
email: jbbru@math.ucdavis.edu}, B.
NACHTERGAELE\footnote{Department of Mathematics, UC Davis -
email: bxn@math.ucdavis.edu} and V.A.
ZAGREBNOV\footnote{Department of Mathematics, UC Davis, -
email: zagreb@math.ucdavis.edu; on leave of absence from\newline
\hspace*{5mm}   Universit\'e de 
la M\'editerran\'ee (Aix-Marseille II) - email : zagrebnov@cpt.univ-mrs.fr}}

\vspace{0.5 cm}

{\bf Department of Mathematics, One Shields Avenue, University of
California , Davis, CA 95616
USA \\[7pt]
Universit\'e de la M\'editerran\'ee and Centre de Physique
Th\'eorique, CNRS-Luminy-Case 907, 13288 Marseille, Cedex 09,
France}

\vspace{1.5cm}

{\bf Abstract}

\end{center}

We study the equilibrium Gibbs states for a Boson gas model,
defined by Bru and Zagrebnov \cite{BruZagrebnov5}, which has two
phase transitions of the Bose condensation type. The two phase
transitions correspond to two distinct mechanisms by which these
condensations can occur. The first ({\it non-conventional}) Bose
condensation is mediated by a zero-mode interaction term in the
Hamiltonian. The second is a transition due to saturation quite
similar to the {\it conventional} Bose-Einstein (BE) condensation
in the ideal Bose gas. Due to repulsive interaction in non-zero
modes the model manifests a generalized type III \textit{i.e.
non-extensive} BE condensation. Our main result is that, as in the
ideal Bose gas, the conventional condensation is accompanied by a
loss of strong equivalence of the canonical and grand canonical
ensembles whereas the non-conventional one, due to the
interaction, does not break the equivalence of ensembles. It is
also interesting to note that the type of (generalized)
condensate, I, II, or III (in the terminology of van den Berg,
Lewis and Pul\'{e} \cite{BergLewis2,Berg1,BergLewisPule}), has no
effect on the equivalence of ensembles. These results are proved
by computing the generating functional of the cyclic
representation of the Canonical Commutation Relation (CCR) for
the corresponding equilibrium Gibbs states.

\vspace{.5cm}


\vspace{\fill}
{\baselineskip=10pt \noindent Copyright \copyright\ 2002 by the
authors. Reproduction of this article in its entirety, by any
means, is permitted for non-commercial purposes.\par }

\renewcommand{\thefootnote}{\fnsymbol{footnote}}

\end{titlepage}

\setcounter{footnote}{0}
\renewcommand{\thefootnote}{\arabic{footnote}}

\Section{Introduction and setup of the problem\label{section BZthermo}}

In recent years, the phenomenon of Bose condensation, described
first by Einstein in 1925 \cite{Einstein}, has become an active
area of research, both experimentally and theoretically. An
example is the existence of a new kind of condensation which was
recently theoretically discovered by an analysis of the
thermodynamic behaviour of the Bogoliubov Weakly Imperfect Bose
Gas
\cite{BruZagrebnov1}-\cite{BruZagrebnov8}
or of some specific Bose systems with diagonal interactions
\cite{BruZagrebnov5,BruZagrebnov7}. This new Bose condensation,
denoted as \textit{non-conventional} Bose condensation, is in
fact induced by a mechanism of \textit{interaction} whereas the
\textit{conventional} one, i.e. the Bose-Einstein (BE)
condensation, appears by a phenomenon of \textit{saturation},
i.e. by the existence only of a \textit{bounded} critical density
\cite{ZiffUhlenbeckKac}-\cite{MichoelVerbeure}.
In fact, the Bose condensation occurring in the
Huang-Yang-Luttinger model and in the, so-called, Full Diagonal
Model, studied in great detail in
\cite{BergLewisPule1}-\cite{DorlasLewisPule2},
should also be considered as examples of the
\textit{non-conventional} type, since in the both cases it is due
to the interaction in those models.

The analysis of the effect of the \textit{conventional} BE
condensation on the equilibrium states was initially worked out
by Araki and Woods in the case of the Perfect Bose Gas (PBG)
\cite{ArakiWoods1}, and further refined in
\cite{Cannon1}-\cite{Lewis2}. A well-known model that exhibits
{\it non-conventional} condensation is the Bogoliubov model
\cite{BruZagrebnov1}-\cite{BruZagrebnov6}, \cite{BruZagrebnov8}.
As a complete and rigorous analysis of the Gibbs states of the
Bogoliubov model is beyond the reach of current techniques, we
propose to analyze the effect of the \textit{non-conventional}
Bose condensation on the Gibbs states in the simpler model
defined in \cite{BruZagrebnov5}, see (\ref{diagmodel1}), in which
the both kinds of Bose condensation occur.

The model we consider is a system of spinless bosons of mass $m$
enclosed in a cubic box $\Lambda \subset \mathbb{R}^{d}$ of volume
$V=\left| \Lambda \right| =L^{d}$ centered at the origin with a
Hamiltonian of the form
\begin{equation}
H_{\Lambda }^{I}=T_{\Lambda }+U_{\Lambda }^{0}+U_{\Lambda
}^{I}=H_{\Lambda }^{0}+U_{\Lambda }^{I},  \label{diagmodel1}
\end{equation}
with
\begin{equation}
\begin{array}{l}
T_{\Lambda }=\stackunder{k\in \Lambda ^{*}\backslash \left\{
0\right\} }{ \dsum }\varepsilon _{k}a_{k}^{*}a_{k},\text{
}\varepsilon _{k}=\hbar
^{2}k^{2}/2m,\text{ for all }k\neq 0, \\
\\
U_{\Lambda }^{0}=\varepsilon _{0}a_{0}^{*}a_{0}+\dfrac{g_{0}}{2V}%
a_{0}^{*}a_{0}^{*}a_{0}a_{0},\text{ }\varepsilon _{0}<0,
\text{ }g_{0}>0, \\
\\
U_{\Lambda }^{I}=\dfrac{1}{2V}\stackunder{k\in \Lambda
^{*}\backslash \left\{ 0\right\} }{\dsum
}g_{k}a_{k}^{*}a_{k}^{*}a_{k}a_{k}, \text{ } g_{+}\geq g_{k}\geq
g_{-}>0.
\end{array}
\label{diagmodel2}
\end{equation}
The sums run over the set
\[
\Lambda ^{*}=\left\{ k\in \mathbb{R}^{d}:\text{ }k_{\alpha
}=\frac{2\pi n_{\alpha }}{L}\text{, }n_{\alpha }=0,\pm 1,\pm
2,...\text{, }\alpha =1,2,...,d\right\} ,
\]
i.e., we consider {\it periodic boundary conditions} on $\partial
\Lambda$. We denote the corresponding one-particle Hilbert space
by $L^{2}\left( \Lambda \right)$. Here $a_{k}^{\#}=\left\{
a_{k}\text{ or }a_{k}^{*}\right\} $ are the usual boson creation
and annihilation operators for the one-particle state $\psi
_{k}\left( x\right) =V^{-\frac{1}{2}}e^{ikx},k\in \Lambda ^{*}$,
$x\in \Lambda $, acting on the boson Fock space
$\mathcal{F}_{\Lambda }^{B}\equiv \mathcal{F}^{B}\left(
L^{2}\left( \Lambda \right) \right) $ over $L^{2}\left( \Lambda
\right)$:
\begin{equation}
\mathcal{F}_{\Lambda }^{B}\equiv
\stackunder{n=0}{\stackrel{+\infty }{\oplus
}}\mathcal{H}_{B}^{\left( n\right) },  \label{definition de
espace de fock}
\end{equation}
where
\begin{equation}
\mathcal{H}_{B}^{\left( n\right) }\equiv \left( L^{2}\left(
\Lambda ^{n}\right) \right) _{\text{symm}} \label{definition de
Hilbert space for n part}
\end{equation}
is the \textit{symmetrized} $n$-particle Hilbert spaces
appropriate for bosons, and $\mathcal{H}_{B}^{\left( 0\right)
}=\mathbb{C}$. We denote by
\[
A^{\left( n\right) }\equiv A\text{ }\lceil
\mathcal{H}_{B}^{\left( n\right) }
\]
the restriction of an operator $A$ acting on the boson Fock
space $\mathcal{F%
}_{\Lambda }^{B}$ to $\mathcal{H}_{B}^{\left( n\right) }.$

This Bose model was introduced and studied in
\cite{BruZagrebnov5}. Clearly, it can be considered as a
perturbation of the kinetic-energy $T_{\Lambda }$ with diagonal
interactions in modes $k=0$ ($U_{\Lambda }^{0}$) and $k\neq 0$
($U_{\Lambda }^{I}$).

The main interest of this model is that it exhibits {\it two}
phase transitions accompanied by the formation of
\textit{non-conventional} and \textit{conventional} Bose
condensation. The first is due to the \textit{negative} effective
excitation energy $\varepsilon _{0}<0$, which leads to a
macroscopic occupation of the zero-mode in some interval of
negative chemical potentials. The  second is a conventional
condensation due to \textit{saturation}. Notice that the second
repulsive term ($g_{0}>0$) in $U_{\Lambda }^{0}$ prevents the
Bose gas from collapse, i.e., it keeps the particle density
finite.

We now summarize the main results of \cite{BruZagrebnov5}, where
it was shown in detail that the model $H_{\Lambda }^{I}$
(\ref{diagmodel1}) displays a two-stage Bose condensation. Let
$\mu $ and $\theta =\beta ^{-1}$ denote the chemical potential
and temperature, respectively. Furthermore
\[
N_{\Lambda }=\stackunder{k\in \Lambda ^{*}}{\sum }N_{k}\equiv \stackunder{%
k\in \Lambda ^{*}}{\sum }a_{k}^{*}a_{k}
\]
is the particle-number operator and $\left\langle -\right\rangle
_{H_{\Lambda }^{I}}\left( \beta ,\mu \right) $ represents the
grand-canonical Gibbs state for the Hamiltonian $H_{\Lambda
}^{I}$. Define
\begin{equation}
\rho ^{P}\left( \beta ,\mu \right) =\frac{1}{\left( 2\pi \right) ^{d}}%
\stackunder{\mathbb{R}^{d}}{\int }\left( e^{\beta \left( \varepsilon
_{k}-\mu \right) }-1\right) ^{-1}dk  \label{densite du PBG}
\end{equation}
and
\begin{equation}
\rho _{c}^{P}\left( \beta \right) \equiv \text{ }\stackunder{%
\mu <0}{\sup }\rho ^{P}\left( \beta ,\mu \right) =\frac{1}{\left(
2\pi \right) ^{d}}\stackunder{\mathbb{R}^{d}}{\int }\left( e^{\beta
\varepsilon _{k}}-1\right) ^{-1}dk<+\infty ,\text{ }d
> 2. \label{densite critique du PBG}
\end{equation}
$\rho ^{P}$ and $\rho _{c}^{P}$ are the density and the critical
density of the PBG, respectively. The following results are
proved in \cite{BruZagrebnov5}:

\begin{itemize}
\item  The model has well-behaved thermodynamics, i.e., the pressure exists,
for the temperature $\theta =\beta ^{-1}\geq 0$, and chemical
potential $\mu \leq 0$. We denote this domain by $Q=\left\{
\left( \theta ,\mu \right) :\theta \geq 0,\text{ }\mu \leq
0\right\}$. Notice that the same is valid for any finite
$\varepsilon_0 \in \mathbb{R}^{1}$.

\item  There is no condensation for $\mu \leq \varepsilon _{0}$, but
condensation occurs for $\varepsilon _{0}<\mu \leq 0$. More
precisely, one has a macroscopic occupation of the $k=0$ mode,
given by
\begin{equation}
\rho _{0}^{I}\left(\beta,\mu \right) \equiv \text{ }
\stackunder{\Lambda }{\lim }\left\langle
\frac{N_{0}}{V}\right\rangle _{H_{\Lambda }^{I}}\left( \beta ,\mu
\right) =\max \left\{ 0,\frac{\mu -\varepsilon
_{0}}{g_{0}}\right\} ,  \label{Generating0}
\end{equation}
i.e., there is Bose condensation due to the instability implied
by the negative excitation energy $\varepsilon _{0}<0$, thought
of as being induced by an interaction mechanism
(\textit{non-conventional} condensation).

\item  For $d > 2$, the non-conventional Bose condensate density $\rho
_{0}^{I}\left( \beta ,\mu \right) $ (\ref{Generating0}) and the
total particle density
\begin{equation}
\rho ^{I}\left( \beta ,\mu \right) \equiv \text{
}\stackunder{\Lambda }{\lim }\left\langle \frac{N_{\Lambda
}}{V}\right\rangle _{H_{\Lambda }^{I}}\left( \beta ,\mu \right)
=\rho ^{P}\left( \beta ,\mu \right) +\rho _{0}^{I}\left( \beta
,\mu \right)  \label{Generating01}
\end{equation}
attain their maxima at $\mu =0$. For densities exceeding a
critical value,
\begin{equation}
\rho >\rho _{c}^{I}\left( \beta \right) \equiv \text{
}\stackunder{\mu \leq 0}{\sup }\rho ^{I}\left( \beta ,\mu \right)
=\stackunder{\mu \rightarrow 0^{-}}{\lim }\rho ^{I}\left( \beta
,\mu \right) =\rho _{c}^{P}\left( \beta \right) -\frac{\varepsilon
_{0}}{g_{0}}<+\infty ,  \label{Generating1}
\end{equation}
the $H_{\Lambda }^{I}$ model (\ref{diagmodel1}) manifests a
generalized type III (i.e., {\it non-extensive}) conventional BE
condensation:
\begin{equation}\label{Generating2}
\widetilde{\rho }_{0}^{I}\left( \beta ,\rho \right) \equiv \text{ }%
\stackunder{\delta \rightarrow 0^{+}}{\lim }\stackunder{\Lambda }{\lim }%
\frac{1}{V}\stackunder{\left\{ k\in \Lambda ^{*}:\text{
}0<\left\| k\right\| <\delta \right\} }{\sum }\left\langle
N_{k}\right\rangle _{H_{\Lambda }^{I}}
=\left\{
\begin{array}{ll}
0 &\text{ for }\rho \leq\rho _{c}^{I}\left( \beta \right) \\
\rho -\rho
_{c}^{I}\left( \beta \right) &\text{ for }
\rho > \rho _{c}^{I}\left( \beta \right)
\end{array}
\right. .
\end{equation}
The term {\it non-extensive} refers to the fact that no single mode 
has a macroscopic occupation of particles.
For $\varepsilon _{0}<0$ this conventional condensate {\it
coexists} at $\mu=0$ with the non-conventional condensate $\rho
_{0}^{I}\left( \beta ,\mu=0\right) $ in the mode $k=0.$ Notice
that the conventional BE condensation (\ref{Generating2}) appears
in spite of the \textit{repulsive} interaction $U_{\Lambda }^{I}$
(\ref{diagmodel2}) between bosons in modes $k\neq 0.$  But it is
because of this repulsion that the condensation is non-extensive.

\item  Remark that formula
(\ref{Generating0}) is also valid for $\varepsilon _{0}\geq 0$.
In this case one gets $\rho _{0}^{I}\left(\beta,\mu \right)
\equiv 0  $, i.e., the model (\ref{diagmodel1}) manifests, for
$d>2$, only the non-extensive conventional condensation
(\ref{Generating2}).
\end{itemize}
More details about the non-extensive BE condensation one finds in
Appendix A of the present paper. These results are an extension
of those of \cite{BruZagrebnov5}. They are indispensable for
calculation of the generating functional for the model
(\ref{diagmodel1}).

We conclude this introduction with a few remarks. The first
concerns the effect of the repulsive term $U_{\Lambda }^{I}$ in
(\ref{diagmodel2}). It is known that for $g_{-}>0$, this
interaction converts the \textit{conventional} condensation from
\textit{type I} (macroscopic occupation of bounded number of
modes $k\neq 0$), e.g., a single mode $k=0$ such as occurs in the
PBG , $\varepsilon_0 = g_{0} = g_{+} = 0$), into one of
\textit{type III } (no macroscopic occupation of a single mode,
accumulation of a finite fraction of the particles in an
infinitesimal interval near $k=0$)
\cite{Zagrebnov,BruZagrebnov5,MichoelVerbeure}. The simplest
example corresponds to the PBG ($\varepsilon_0 = g_{0} = g_{+} =
0$ )in an \textit{isotropically} dilated container, when the
macroscopic occupation of
the single mode $k=0$ is transformed by the pure repulsive
interaction, $g_{0}>0, g_{-}>0, \varepsilon_0 = 0$, into a
non-extensive BE condensation \cite{Zagrebnov}. We stress here the
\textit{isotropic} shape of the container, since the \textit{conventional}
condensation is so subtle matter that the PBG itself manifests the
non-extensive BE condensation, if for example, dilated rectangular
box has a highly inisotropic growth rates for the edges   \cite{Zagrebnov,BergLewis1,BergLewisPule,BergLewisLunn}.

Our second remark concerns the \textit{dimension} dependence of
the phase transitions. In contrast to the \textit{conventional}
condensation caused by saturation for $d>2$, the
\textit{non-conventional} condensation (\ref{Generating0}) is due
to interaction, and it exists for all dimensions, including
$d=1$, and $2$. This is another indication that the simplified
model (\ref{diagmodel1}) with diagonal interaction is similar to
the Bogoliubov Gas \cite{BruZagrebnov6}-\cite{BruZagrebnov8}.
Moreover, in contrast to conventional BE condensations, the
non-conventional condensation may emerge as a first-order phase
transitions: the Bose condensate density appears discontinuously,
see for example the thermodynamic behaviour of the Bogoliubov gas
\cite{BruZagrebnov1}-\cite{BruZagrebnov4}, \cite{BruZagrebnov8},
or the Huang-Yang-Luttinger model, see
\cite{BergLewisPule1,BergDorlasLewisPule1} and \cite{HYL}. This
transition in the simplified model studied here is continuous
however, see (\ref{Generating0}).

The bulk of the present paper is devoted to the study of the
Gibbs states of the model (\ref{diagmodel1}). In particular we
shall calculate the generating functional of the cyclic
representation of the Canonical Commutation Relations (CCR) for
the Gibbs states of the model (\ref{diagmodel1}), a method
introduced in 1963 by Araki and Woods \cite{ArakiWoods1}. {From}
the generating functional it is then straightforward the read off
properties such as the breakdown of the strong equivalence of
ensembles as was done in \cite
{Cannon1}-\cite{Lewis2}, \cite{BrattelliRobinson}. We shall see that
the two phase transitions have their distinct effects on the
generating functional.

Before we embark on the actual calculation we present, in Section
\ref {chapitreetat}, the relevant known properties of the
generating functional of the cyclic representation of the CCR for
the Gibbs states of the PBG \cite {BrattelliRobinson}. In Section
\ref{etatmodelBZ} we calculate in the thermodynamic limit the
grand-canonical generating functional for the Gibbs state $
\left\langle -\right\rangle _{H_{\Lambda }^{I}}\left( \beta ,\mu
\right) $ associated with the model (\ref{diagmodel1}) for a
fixed chemical potential $\mu <  0 $, or a fixed density $\rho <
\rho _{c}^{I}\left( \beta \right) $.  In the next Section
\ref{Section 4} we determine the generating functional for a
fixed particle density $\rho \geq\rho _{c}^{I}\left( \beta
\right) $ (\ref{Generating1}). In Section \ref {Section 5} we
summarize our conclusions and formulate some tentative
generalizations. Some technical results are collected in the
Appendix A.

\Section{Generating functionals\label{chapitreetat}}

The purpose of this section is to review the characterization of
(Gibbs) states of a Bose system by their generating functional, a
method originally introduced by Araki and Woods in the case of
the PBG \cite{ArakiWoods1}. For each Gibbs state there is a
representation of the Canonical Commutation Relations (CCR) given
by the GNS construction. For a complete description see
\cite{BrattelliRobinson}, and also
\cite{Cannon1}-\cite{Lewis2} for a detailed analysis of the
PBG Gibbs state. Here, we only present a quick overview.

Let $M$ be a complex pre-Hilbert space with the corresponding
scalar product $\left( .,.\right) _{M}$. We consider a
representation of the CCR over $M$ given by a map $h\mapsto
W\left( h\right)$ from $M$ to a space $U\left( \mathcal{H}\right)
$ of unitary operators on a Hilbert space $\mathcal{H}$ satisfying
\begin{equation}
W\left( h_{1}\right) W\left( h_{2}\right) =\exp \left\{ -\frac{i}{2}\func{Im}%
\left( h_{1},h_{2}\right) _{M}\right\} W\left( h_{1}+h_{2}\right)
, \label{CCR5}
\end{equation}
and such that the map $\lambda \mapsto W\left( \lambda h\right) $ from $\mathbb{%
R}$ to $U\left( \mathcal{H}\right) $ is strongly continuous. By
Stone's theorem \cite{BrattelliRobinson}, the continuity implies
the existence of self-adjoint operators $R\left( h\right) $ such
that
\begin{equation}
W\left( h\right) =\exp \left\{ iR\left( h\right) \right\} .
\label{CCR2}
\end{equation}
The $R(h)$ are called the \emph{field operators} and can be
interpreted as the random variables of a non-commutative
probability theory, since by (\ref{CCR5}) one gets
\begin{equation}
\left[R(h_1),R(h_2)\right]=i\func{Im}\left( h_{1},h_{2}\right)
_{M}.\label{RR}
\end{equation}
Note that the map $h\rightarrow R\left( h\right) $ is a linear
over $\mathbb{R}$. 
For $h\in M$, we
can now define the \emph{creation} and \emph{annihilation}
operators $a^{*}\left( h\right)$ and $a\left( h\right) \equiv
(a^{*}\left( h\right))^{*} $ by
\begin{equation}
a^{*}\left( h\right) \equiv \frac{1}{\sqrt{2}}\left\{ R\left(
h\right)
-iR\left( ih\right) \right\} ,\text{ }a\left( h\right) \equiv \frac{1}{\sqrt{%
2}}\left\{ R\left( h\right) +iR\left( ih\right) \right\} .
\label{CCR3}
\end{equation}

A representation of the CCR is called \emph{cyclic} if there is a vector $%
\Omega $ in $\mathcal{H}$ such that the set $\left\{ W\left(
h\right) \Omega \right\} _{h\in M}$ is dense in $\mathcal{H}$.
Such $\Omega $ is called a cyclic vector. It can be shown that,
for every {\it regular} Gibbs state $\langle \cdot \rangle $,
there is unique (up to unitary equivalence) representation of the
CCR with cyclic vector $\Omega $ such that
\[
\left\langle \exp \left\{ iR\left( h\right) \right\}
\right\rangle =(\Omega ,W(h)\Omega) _{\mathcal{H}}.
\]
The \emph{generating functional} of the representation is defined
by
\begin{equation}
\mathbb{E}\left( h\right) \equiv \left( \Omega ,W\left( h\right)
\Omega \right) _{\mathcal{H}},\text{ }h\in M.
\label{CCR4bisCCR4bis}
\end{equation}
The generating functional plays the same r\^{o}le for a state on
the CCR algebra, as the characteristic function for probability
distribution, see \cite{BrattelliRobinson}.
\begin{theorem}\label{Araki-Segal}(Araki-Segal)
Let $\mathbb{E}$ be the generating functional of a cyclic
representation of the CCR over $M$. Then it satisfies :
\newline $\left( i\right) $ $\mathbb{E}\left( 0\right) =1;$\newline
$\left( ii\right) $ for any finite set $\left\{ c_{j}\in \mathbb{C};\text{ }%
h_{j}\in M\right\} $, one has
\[
\stackunder{l,s=1}{\stackrel{n}{\sum }}\mathbb{E}\left(
h_{l}-h_{s}\right) \exp \left\{ \frac{i}{2}\func{Im}\left(
h_{l},h_{s}\right) _{M}\right\} \overline{ c_{l}}c_{s}\geq 0;
\]
$\left( iii\right) $ for $h\in M$, the map $\lambda \rightarrow \mathbb{E}%
\left( \lambda h\right) $ from $\mathbb{C}$ to $\mathbb{R}$ is
continuous.
\newline
Conversely, any generating functional
$\mathbb{E}:M\rightarrow \mathbb{C}$ satisfying $\left( i\right) $,
$\left( ii\right) $ and $\left( iii\right) $ is a generating
functional of a cyclic representation of the CCR.
\end{theorem}
Our concrete setup will be as follows. For a (sufficiently
regular) \textit{ finite} volume, $\Lambda \subset \mathbb{R}^{d}$,
the grand canonical Gibbs state $\left\langle \cdot \right\rangle
_{\Lambda }\left( \beta ,\mu \right) $, is defined on the set of
bounded operators acting on the boson Fock space
$\mathcal{F}_{\Lambda }^{B}\equiv \mathcal{F}^{B}\left(
L^{2}\left( \Lambda \right) \right) $ over $L^{2}\left( \Lambda
\right) $, see (\ref{definition de espace de fock}). In order to
analyze the state $\left\langle \cdot \right\rangle _{\Lambda
}\left( \beta ,\mu \right) $, we use the Fock representation
$W^{\mathcal{F}_{\Lambda }^{B}}$of the CCR over the pre-Hilbert
space $M=\mathcal{D}_{\Lambda }$ (the space of the
$C_{0}^{\infty}(\Lambda)$-functions with compact supports
contained in $\Lambda$). Its generating functional
(\ref{CCR4bisCCR4bis}) is equal to $\mathbb{E}_{\mathcal{F}_{\Lambda
}^{B}}(h) =e^{-\frac{1}{4}\left\| h\right\|^{2}}$, where cyclic
vector $\Omega$ is vacuum in $\mathcal{H}=\mathcal{F}_{\Lambda
}^{B}$: $a(h)\Omega = 0$ for any $h\in \mathcal{D}_{\Lambda }$.
Since $\mathcal{D}_{\Lambda }$ is dense in  $L^{2}\left( \Lambda
\right)$, one can extend $W^{\mathcal{F}_{\Lambda }^{B}}$ to the
later. We shall calculate the generating functional
\begin{equation}
\mathbb{E}_{\Lambda }\left( \beta ,\mu ;h\right) \equiv \left\langle W^{%
\mathcal{F}_{\Lambda }^{B}}\left( h\right) \right\rangle
_{\Lambda }\left( \beta ,\mu \right) ,\text{ }h\in
\mathcal{D}_{\Lambda }, \label{CCR4bis}
\end{equation}
and study its thermodynamic limit ($\Lambda \uparrow
\mathbb{R}^{d}$).

\Section{Gibbs state and non-conventional condensation\label{etatmodelBZ}}

Recall from Section \ref{section BZthermo} that condensation in
the exactly solvable model $H_{\Lambda }^{I}$ (\ref{diagmodel1})
occurs in two stages: for intermediate densities $\rho < \rho
_{c}^{I}\left( \beta \right)$, i.e., for negative chemical
potentials $\varepsilon _{0}<\mu < 0$, one has only
\textit{non-conventional} Bose condensation in the $k=0$ mode due
to the diagonal perturbation $U_{\Lambda }^{0}$ (\ref
{diagmodel2}) of the PBG (cf. (\ref{Generating0})), whereas for
large densities $\rho \geq\rho _{c}^{I}\left( \beta \right)$
($\mu=0$), this condensate coexists with conventional (type III)
generalized BE condensation corresponding to the standard
mechanism of saturation, see (\ref{Generating1}) and (\ref
{Generating2}).

In this section we study the influence on the corresponding Gibbs
state of the first stage of condensation: the
\textit{non-conventional} one (\ref {Generating0}) which appears
for a fixed chemical potential $\varepsilon _{0}<\mu \leq 0$.
Following \cite{ArakiWoods1}-\cite{Lewis2}, we use
the Fock representations of the CCR \cite{BrattelliRobinson} over
the space $\mathcal{D}_{\Lambda }$ of $C^{\infty}$-smooth functions with
compact support contained in $\Lambda $ (Section \ref
{chapitreetat}) and we define by
\begin{equation}
\mathbb{E}_{\Lambda }^{I}\left( \beta ,\mu ;h\right) \equiv \left\langle W^{%
\mathcal{F}_{\Lambda }^{B}}\left( h\right) \right\rangle
_{H_{\Lambda }^{I}}\left( \beta ,\mu \right) ,\text{ }h\in
\mathcal{D}_{\Lambda }, \label{CCR21}
\end{equation}
the grand-canonical generating functional of the model
(\ref{diagmodel1}):
\begin{equation}
H_{\Lambda }^{I}=\stackunder{k\in \Lambda ^{*}}{\sum }\left\{
\varepsilon_{k}a_{k}^{*}a_{k}+\frac{g_{k}}{2V}a_{k}^{*^{2}}a_{k}^{2}\right\} 
=\stackunder{k\in \Lambda ^{*}}{\sum }H_{k}^{I},\text{
}\varepsilon _{k\neq 0}=\hbar ^{2}k^{2}/2m\geq 0\text{,
}\varepsilon _{0}<0.  \label{CCR20}
\end{equation}
Here the operators $W^{^{\mathcal{F}_{\Lambda }^{B}}}\left(
h\right)$ for $h\in
\mathcal{D}_{\Lambda }$, are defined by (\ref{CCR2})-(\ref{CCR3}).

Note that the boson Fock space $\mathcal{F}_{\Lambda }^{B}$
(\ref{definition de espace de fock}) is isomorphic to the tensor
product: $\mathcal{F}_{\Lambda }^{B} \approx \stackunder{k\in
\Lambda ^{*}}{\otimes }\mathcal{F}_{k}^{B}$, where $\mathcal{F}%
_{k}^{B}\equiv \mathcal{F}^{B}\left( \mathcal{H}_{k}\right) $ is
the boson Fock space constructed on the one-dimensional Hilbert
space
\begin{equation}
\mathcal{H}_{k}=\left\{ \lambda e^{ikx}\right\} _{\lambda \in
\mathbb{C}}. \label{CCR5bisbis}
\end{equation}
Then using the Fourier decomposition
\begin{equation}
\begin{array}{l}
a^{*}\left( h\right) =\stackunder{\Lambda }{\dint }dxh\left(
x\right)
a^{*}\left( x\right) =\dfrac{1}{\sqrt{V}}\stackunder{k\in \Lambda ^{*}}{%
\dsum }\left( e^{ikx},h\right) _{L^{2}\left( \Lambda \right)
}a^{*}\left(
\psi _{k}\right) \equiv \dfrac{1}{\sqrt{V}}\stackunder{k\in
\Lambda ^{*}}{\dsum }%
h_{k}a_{k}^{*}, \\
a\left( h\right) =\stackunder{\Lambda }{\dint }dx\overline{h\left(
x\right) }a\left( x\right) =\dfrac{1}{\sqrt{V}}\stackunder{k\in \Lambda ^{*}%
}{\dsum }\left( h,e^{ikx}\right) _{L^{2}\left( \Lambda \right)
}a\left( \psi _{k}\right) \equiv
\dfrac{1}{\sqrt{V}}\stackunder{k\in \Lambda ^{*}}{\dsum }
\overline{h_{k}}a_{k},
\end{array}
\label{CCR5bis}
\end{equation}
we can write the generating functional $\mathbb{E}_{\Lambda
}^{I}\left( \beta ,\mu ;h\right) $ (\ref{CCR21}) in the following
form:
\begin{eqnarray}
\mathbb{E}_{\Lambda }^{I}\left( \beta ,\mu ;h\right)
&=& \stackunder{k\in
\Lambda ^{*}}{\prod }\left\langle e^{\frac{i}{
\sqrt{2V}}\left( \overline{h_{k}}a_{k}+h_{k}a_{k}^{*}\right)
}\right\rangle _{H_{\Lambda}^{I}}(\beta,\mu)
\nonumber \\
&=&\stackunder{k\in \Lambda ^{*}}{\prod
}\frac{\Tr_{\mathcal{F}_{k}^{B}}\left( e^{-\beta
H_{k}^{I}\left( \mu \right) }e^{\frac{i}{\sqrt{2V}}\left( \overline{h_{k}}%
a_{k}+h_{k}a_{k}^{*}\right) }\right)
}{\Tr_{\mathcal{F}_{k}^{B}}\left(
e^{-\beta H_{k}^{I}\left( \mu \right) }\right) }  \nonumber \\
&=&\left\langle e^{\frac{i}{\sqrt{2V}}\left( \overline{h_{0}}%
a_{0}+h_{0}a_{0}^{*}\right) }\right\rangle
_{H_{0}^{I}}\stackunder{k\in
\Lambda ^{*}\backslash \left\{ 0\right\} }{\prod }\left\langle e^{\frac{i}{%
\sqrt{2V}}\left( \overline{h_{k}}a_{k}+h_{k}a_{k}^{*}\right)
}\right\rangle _{H_{k}^{I}}.  \label{CCR22}
\end{eqnarray}
Next, we study the two factors corresponding to cases $k=0$ and
$k\neq 0$ separately. Denote by $\mathcal{D}= \bigcup_{\Lambda
\subset \mathbb{R}^d}\mathcal{D}_{\Lambda}$ the space of $C^{\infty}$-smooth
functions on $\mathbb{R}^d$ having compact support, and by
$\widehat{h}_k \equiv \left( e^{ikx},h\right) _{L^{2}\left(
\mathbb{R}^d \right)}$, $k \in \mathbb{R}^d$.

\begin{theorem}\label{CCRth9}
Let $\varepsilon _{0} \in \mathbb{R}^1$ and $g_{0}>0$. Suppose that
$0\leq g_{-}\leq g_{k}\leq g_{+}$ for $k\in \Lambda ^{*}\backslash
\left\{ 0\right\}$. Then for $\mu <0$ and any $h$ in the space
$\mathcal{D}$ one gets that:
\newline
$\left( i\right) \ $ for the mode $k=0$
\begin{equation}
\stackunder{\Lambda }{\lim }\left\langle
e^{\frac{i}{\sqrt{2V}}\left(
\overline{h_{0}}a_{0}+h_{0}a_{0}^{*}\right) }\right\rangle
_{H_{0}^{I}}=J_{0}\left( \sqrt{2\rho _{0}^{I}\left( \beta ,\mu
\right) } \left| \widehat{h}_{0}\right| \right) , \label{CCR17}
\end{equation}
where the non-conventional Bose-condensate density $\rho
_{0}^{I}\left( \beta ,\mu \right) \ $ is defined by
(\ref{Generating0});
\newline
$\left( ii\right) \ $ for the second factor in (\ref{CCR22}) we
have
\begin{equation}
\stackunder{\Lambda }{\lim }\stackunder{k\in \Lambda
^{*}\backslash \left\{
0\right\} }{\prod }\left\langle e^{\frac{i}{\sqrt{2V}}
\left( \overline{h_{k}}%
a_{k}+h_{k}a_{k}^{*}\right) }\right\rangle _{H_{k}^{I}}=
\exp \left\{ -\frac{1%
}{4}\left\| h\right\| ^{2}-\frac{1}{2}A_{\beta ,\mu }\left(
h,h\right) \right\} ,  \label{CCR23}
\end{equation}
where the sesquilinear form $A_{\beta ,\mu }\left( u , v \right)$, for
$u,v \in \mathcal{D}$, is defined by
\begin{equation}
A_{\beta ,\mu }\left( u , v\right) =
\frac{1}{\left( 2\pi \right) ^{d}}\stackunder{\mathbb{R}^{d}}{%
\int }\dfrac{\overline {\widehat{u}_{k}} \
\widehat{v}_{k}}{e^{\beta \left( \varepsilon _{k}-\mu \right)
}-1}dk . \label{CCR11bis}
\end{equation}
\end{theorem}
\noindent \textit{Proof.} \ $\left( i\right) $ Let $\left\{ \psi
_{n}\right\} _{n\geq 0}\subset \mathcal{F}_{0}^{B}$ be an
orthonormal base of eigenvectors of the operator $a_{0}^{*}a_{0}$:
\[
a_{0}^{*}a_{0}\psi _{n}=n\psi _{n}.
\]
Then one gets
\begin{equation}
X\equiv \Tr_{\mathcal{F}_{0}^{B}}\left( e^{-\beta \left(
H_{0}^{I}-\mu
N_{0}\right) }e^{\frac{i}{\sqrt{2V}}\left( \overline{h_{0}}%
a_{0}+h_{0}a_{0}^{*}\right) }\right) =\stackunder{n=0}{\stackrel{+\infty }{%
\sum }}e^{-\beta \left[ \left( \varepsilon _{0}-\mu
-\frac{g_{0}}{2V}\right)
n+\frac{g_{0}}{2V}n^{2}\right] }\left( \psi _{n},e^{\frac{i}{\sqrt{2V}}%
\left( \overline{h_{0}}a_{0}+h_{0}a_{0}^{*}\right) }\psi _{n}\right) _{%
\mathcal{F}_{\Lambda }^{B}}.  \label{Generating3}
\end{equation}
By the Baker-Campbell-Hausdorff formula:
\begin{equation}
e^{A+B}=e^{A}e^{B}e^{-\frac{1}{2}\left[ A,B\right] },\text{ if
}\left[ A,\left[ A,B\right] \right] =\left[ B,\left[ A,B\right]
\right] =0, \label{CCR6bis}
\end{equation}
we obtain that
\begin{equation}
\exp \left\{ \frac{i}{\sqrt{2V}}\left( \overline{h_{0}}a_{0}+h_{0}a_{0}^{*}%
\right) \right\} =e^{-\frac{1}{4V}\left| h_{0}\right| ^{2}}\exp
\left\{
\frac{i}{\sqrt{2V}}h_{0}a_{0}^{*}\right\} \exp \left\{ \frac{i}{\sqrt{2V}}%
\overline{h_{0}}a_{0}\right\} .  \label{CCR17bis}
\end{equation}
Therefore, since
\begin{eqnarray*}
a_{0}\psi _{n} &=&\sqrt{n}\psi _{n-1}, \\
a_{0}^{*}\psi _{n} &=&\sqrt{n+1}\psi _{n+1}, \\
\left(\psi _{n},\psi _{n'}\right)_{\mathcal{F}_{\Lambda }^{B}}
&=& \delta_{n,n'},
\end{eqnarray*}
by (\ref{CCR17bis}) the trace (\ref{Generating3}) equals:
\begin{eqnarray}
X &=&e^{-\frac{1}{4V}\left| h_{0}\right| ^{2}}\stackunder{n=0}{\stackrel{%
+\infty }{\sum }}e^{-\beta \left[ \left( \varepsilon _{0}-\mu -\frac{g_{0}}{%
2V}\right) n+\frac{g_{0}}{2V}n^{2}\right] }\left\| e^{\frac{i}{\sqrt{2V}}%
\overline{h_{0}}a_{0}}\psi _{n}\right\| ^{2}  \nonumber \\
&=&e^{-\frac{1}{4V}\left| h_{0}\right| ^{2}}\stackunder{n=0}{\stackrel{%
+\infty }{\sum }}e^{-\beta \left[ \left( \varepsilon _{0}-\mu -\frac{g_{0}}{%
2V}\right) n+\frac{g_{0}}{2V}n^{2}\right] }\stackunder{l=0}{\stackrel{n}{%
\sum }}\left( -\frac{\left| h_{0}\right| ^{2}}{2V}\right) ^{l}\frac{n!}{%
\left( l!\right) ^{2}\left( n-l\right) !}.  \nonumber \\
&&  \label{CCR15}
\end{eqnarray}
Using the Laguerre polynomials
\[
L_{n}\left( z\right) \equiv \text{ }\stackunder{l=0}{\stackrel{n}{\sum }}%
\frac{n!}{\left( l!\right)^{2}(n-l)!} \left( -z\right)^{l}, \ \text{ }n\geq 0,
\]
(\ref{CCR15}) can be rewritten to give
\[
X=e^{-\frac{1}{4V}\left| h_{0}\right| ^{2}}\stackunder{n=0}{\stackrel{%
+\infty }{\sum }}L_{n}\left( \frac{\left| h_{0}\right| ^{2}}{2V}%
\right) e^{-\beta \left[ \left( \varepsilon _{0}-\mu -\frac{g_{0}}{2V}%
\right) n+\frac{g_{0}}{2V}n^{2}\right] }.
\]
Consequently, we obtain
\begin{equation}
\left\langle e^{\frac{i}{\sqrt{2V}}\left( \overline{h_{0}}%
a_{0}+h_{0}a_{0}^{*}\right) }\right\rangle _{H_{0}^{I}}=e^{-\frac{1}{4V}%
\left| h_{0}\right| ^{2}}\frac{\stackunder{n=0}{\stackrel{+\infty
}{\sum }} L_{n}\left( \frac{\left| h_{0}\right|
^{2}}{2n}\frac{n}{V}\right) e^{-\beta
V\left[ \left( \varepsilon _{0}-\mu -\frac{g_{0}}{2V}\right) \frac{n}{V}+%
\frac{g_{0}}{2}\left( \frac{n}{V}\right) ^{2}\right] }}{\stackunder{n=0}{%
\stackrel{+\infty }{\sum }}e^{-\beta V\left[ \left( \varepsilon _{0}-\mu -%
\frac{g_{0}}{2V}\right) \frac{n}{V}+\frac{g_{0}}{2}\left(
\frac{n}{V}\right) ^{2}\right] }}.  \label{CCR16}
\end{equation}
Notice that the probability distributions in (\ref{CCR16}):
\begin{equation}\label{distr}
F_{V}(x)\equiv \frac{\stackunder{0\leq n/V < x}{\stackrel{}{\sum
}}  e^{-\beta
V\left[ \left( \varepsilon _{0}-\mu -\frac{g_{0}}{2V}\right) \frac{n}{V}+%
\frac{g_{0}}{2}\left( \frac{n}{V}\right) ^{2}\right] }}{\stackunder{n=0}{%
\stackrel{+\infty }{\sum }}e^{-\beta V\left[ \left( \varepsilon _{0}-\mu -%
\frac{g_{0}}{2V}\right) \frac{n}{V}+\frac{g_{0}}{2}\left(
\frac{n}{V}\right) ^{2}\right] }}
\end{equation}
satisfies the Laplace \textit{large deviation principle}
\cite{Lewis1,LewisPfister1}:
\begin{equation}\label{LDP}
\stackunder{\Lambda}{\lim }\int_{\mathbb{R}^1}F_{V}(dx)f(x)=
f(\max\{0,\frac{\mu-\varepsilon_0}{g_{0}}\}),
\end{equation}
for any bounded continuous function $f$ on $\mathbb{R}^1$.

The Laguerre polynomials have the property
\begin{equation}\label{CCR16bis}
\stackunder{n\rightarrow +\infty }{\lim }L_{n}\left( z/n\right)
=J_{0}\left( 2\sqrt{z}\right)= \sum_{l=0}^{\infty}\frac{1}
{\left(l!\right)^{2}}\left(-z\right)^{l} , \quad \mbox{for} \;z\in
\mathbb{C},
\end{equation}
as entire analytic functions in $\mathbb{C}$. Here $J_{0}\left(
x\right) $ is the Bessel function of order 0. Using this and
(\ref{LDP}), we find the thermodynamic limit of (\ref{CCR16}) to
be
\[
\stackunder{\Lambda }{\lim }\left\langle
e^{\frac{i}{\sqrt{2V}}\left(
\overline{h_{0}}a_{0}+h_{0}a_{0}^{*}\right) }\right\rangle
_{H_{0}^{I}}=J_{0}\left( \sqrt{2 \rho _{0}^{I}\left( \theta ,\mu
\right) \left| \widehat{h}_{0}\right| ^{2}}\right) ,
\]
for any $h \in \mathcal{D}$, since $h_0=\widehat{h}_{0}$ for
$\Lambda$ sufficiently large. Thus, by definition
(\ref{Generating0}), we deduce (\ref{CCR17}).
\newline
$\left( ii\right) $ Similar to the proof of (\ref{CCR16}) we
get for $k\in \Lambda ^{*}\backslash \left\{ 0\right\} \ $ that
\begin{equation}
\Gamma_k \equiv \left\langle e^{\frac{i}{\sqrt{2V}}\left( \overline{h_{k}}%
a_{k}+h_{k}a_{k}^{*}\right) }\right\rangle _{H_{k}^{I}}=e^{-\frac{1}{4V}%
\left| h_{k}\right| ^{2}}\frac{\stackunder{n=0}{\stackrel{+\infty }{\sum }}%
L_{n}\left( \frac{\left| h_{k}\right| ^{2}}{2V}\right) e^{-\beta
V\left[ \left( \varepsilon _{k}-\mu -\frac{g_{k}}{2V}\right) \frac{n}{V}+%
\frac{g_{k}}{2}\left( \frac{n}{V}\right) ^{2}\right] }}{\stackunder{n=0}{%
\stackrel{+\infty }{\sum }}e^{-\beta V\left[ \left( \varepsilon _{k}-\mu -%
\frac{g_{k}}{2V}\right) \frac{n}{V}+\frac{g_{k}}{2}\left(
\frac{n}{V}\right) ^{2}\right] }}.  \label{CCR25}
\end{equation}
Let
\begin{equation} \label{newCCR25}
\gamma_k \equiv e^{-\frac{1}{4V}
\left| h_{k}\right| ^{2}}\frac{\stackunder{n=0}{\stackrel{+\infty }{\sum }}%
L_{n}\left( \frac{\left| h_{k}\right| ^{2}}{2V}\right) e^{-\beta
\left[ \left( \varepsilon _{k}-\mu -\frac{g_{k}}{2V}\right)n
\right] }} {\stackunder{n=0}{ \stackrel{+\infty }{\sum
}}e^{-\beta \left[ \left( \varepsilon _{k}-\mu - \frac{g_{k}}{2V}
\right) n \right]}}.
\end{equation}
Since
\begin{equation} \label{CCR27bis}
\stackunder{n=0}{\stackrel{+\infty }{\sum }}L_{n}\left( z\right) s^{n}
=\dfrac{1}{1-s}\exp \left\{ -z\dfrac{s}{1-s}\right\} ,
\end{equation}
we readily get that
\begin{equation}
\stackunder{\Lambda }{\lim }\stackunder{k\in \Lambda
^{*}\backslash \left\{
0\right\} }{\prod } \gamma_k =
\exp \left\{ -\frac{1%
}{4}\left\| h\right\| ^{2}-\frac{1}{2}A_{\beta ,\mu }\left(
h,h\right) \right\} ,  \label{newCCR23}
\end{equation}
see (\ref{CCR23}).

To show that one gets the same limit for $\;\; \stackunder{\Lambda }{\lim }
\stackunder{k\in \Lambda
^{*}\backslash \left\{
0\right\} }{\prod } \Gamma_k \;\;$  we define
\begin{equation}\label{2newCCR25}
\Gamma_{k}(t_{k})\equiv e^{-\frac{1}{4V}%
\left| h_{k}\right|^{2}}\frac{\stackunder{n=0}{\stackrel{+\infty }{\sum }}
L_{n}\left( \frac{\left| h_{k}\right|^{2}}{2V}\right) e^{-\beta
V\left[\left( \varepsilon _{k}-\mu -\frac{g_{k}}{2V}\right) \frac{n}{V}+
\frac{t_{k}}{2}(\frac{n}{V})^{2}\right]}}{\stackunder{n=0}{
\stackrel{+\infty }{\sum }}e^{-\beta V\left[ \left( \varepsilon _{k}-\mu -
\frac{g_{k}}{2V}\right) \frac{n}{V}+\frac{t_{k}}{2}(\frac{n}{V})^{2}
\right] }}.
\end{equation}
Therefore, $\Gamma_{k}(t_{k}=0)=\gamma_{k}$ and
$\Gamma_{k}(t_{k}=g_k)=\Gamma_{k}$. Since
$\Gamma_{k}(t_{k})
\in\mathcal{C}^{\infty}(\mathbb{R}^{1}_{+})$  for each $k\in \Lambda
^{*}\backslash \left\{ 0\right\} \ $, then to prove that $\;\;
\stackunder{\Lambda }{\lim }\stackunder{k\in \Lambda
^{*}\backslash \left\{ 0\right\} }{\prod } \Gamma_k \;\;$
coincides with (\ref{newCCR23}) it is sufficient to estimate the
asymptotic behaviour of derivative
\begin{equation}
\partial _{t_{k}}\Gamma _{k}(t_{k}=0)=\frac{\Delta _{k}(V)}{\left\{ 
\stackunder{n=0}{\stackrel{+\infty }{\tsum }}e^{-\beta \left[ \left(
\varepsilon _{k}-\mu -\frac{g_{k}}{2V}\right) n\right] }\right\} ^{2}}e^{-%
\frac{1}{4V}\left| h_{k}\right| ^{2}},  \label{derivGamma}
\end{equation}
for $ V \rightarrow \infty $. Here
\begin{eqnarray}\label{derivGamma1}
\Delta_{k}(V)&=&\stackunder{n=0}{\stackrel{+\infty }{\tsum
}}\left[\{L_{n}\left( \frac{\left|
h_{k}\right|^{2}}{2V}\right)-1\} e^{-\beta \left[ \left(
\varepsilon _{k}-\mu - \frac{g_{k}}{2V}\right)n
\right]}\left(-\frac{\beta
n^{2}}{2V}\right)\right]\stackunder{n=0}{\stackrel{+\infty }{\tsum
}}e^{-\beta \left[ \left( \varepsilon _{k}-\mu -
\frac{g_{k}}{2V}\right) n\right]}
\nonumber\\
&+&\stackunder{n=0}{\stackrel{+\infty }{\tsum
}}\left[\{L_{n}\left( \frac{\left|
h_{k}\right|^{2}}{2V}\right)-1\} e^{-\beta \left[ \left(
\varepsilon _{k}-\mu - \frac{g_{k}}{2V}\right)n \right]}\right]
\stackunder{n=0}{ \stackrel{+\infty }{\tsum }}\left[e^{-\beta
\left[ \left( \varepsilon _{k}-\mu - \frac{g_{k}}{2V}\right)
n\right]}\left(\frac{\beta n^{2}}{2V}\right)\right].
\end{eqnarray}
Since (\ref{CCR16bis}) implies the convergence of derivatives, one
gets the estimate
\begin{equation}\label{derivGamma2}
\left|   \frac{1}{n}L'_{n}(\frac{z}{m})\right|\leq C_{z_0},\;\;
z\in[0,z_0],
\end{equation}
for any $z_0>0$ and $n\leq m$. Therefore, in this domain we have :
\begin{equation}\label{derivGamma3}
\left|L_{n}(\frac{z}{m})-1\right|\leq C_{z_0}n\frac{z}{m}.
\end{equation}
Let $n_{0}(V)\equiv \left[ V^{1-\delta} \right]$ for some $\delta\in (0,1)$.
Here $\left[x \right]$ denotes the integer part of the real $x$.
Then, since $g_{k}\leq g_{+}$, by virtue of (\ref{derivGamma3})
we can find $C_k>0$ such that for any $\mu<0$ one gets the
estimates :
\begin{eqnarray}\label{estimDelta1}
&&\left|\stackunder{n=0}{\stackrel{ n_{0}(V)}{\tsum
}}\left[\{L_{n}\left( \frac{\left|
h_{k}\right|^{2}}{2V}\right)-1\} e^{-\beta \left[ \left(
\varepsilon _{k}-\mu - \frac{g_{k}}{2V}\right)n
\right]}\left(-\frac{\beta
n^{2}}{2V}\right)\right]\stackunder{n=0}{\stackrel{+\infty }{\tsum
}}e^{-\beta \left[ \left( \varepsilon _{k}-\mu -
\frac{g_{k}}{2V}\right) n\right]}\right|
\nonumber\\
&& \leq C_k
\frac{\beta \left|h_{k}\right|^{2} }{4V^{2}}\left[\partial_{y}^{3}f(y=\beta(\varepsilon
_{k}-\mu))\right]f(\beta(\varepsilon _{k}-\mu)),
\end{eqnarray}
and
\begin{eqnarray}
&&\left| \stackunder{n=0}{\stackrel{n_{0}(V)}{\tsum }}\left[ \{L_{n}\left( 
\frac{\left| h_{k}\right| ^{2}}{2V}\right) -1\}e^{-\beta \left[ \left(
\varepsilon _{k}-\mu -\frac{g_{k}}{2V}\right) n\right] }\right] \stackunder{%
n=0}{\stackrel{+\infty }{\tsum }}\left[ e^{-\beta \left[ \left( \varepsilon
_{k}-\mu -\frac{g_{k}}{2V}\right) n\right] }\left( \frac{\beta n^{2}}{2V}%
\right) \right] \right|  \nonumber \\
&\leq &C_{k}\frac{\beta \left| h_{k}\right| ^{2}}{4V^{2}}\bigl[ \partial
_{y}f(y=\beta (\varepsilon _{k}-\mu ))\bigr] \bigl[ \partial
_{y}^{2}f(y=\beta (\varepsilon _{k}-\mu ))\bigr] ,  \label{estimDelta2}
\end{eqnarray}
where $f(\beta (\varepsilon _{k}-\mu ))=(1-e^{-\beta (\varepsilon _{k}-\mu
)})^{-1}$.

On the other hand, for large $n$ the Laguerre polynomials have the
following asymptotics :
\begin{equation}\label{asympLag1}
L_{n}(x)=\frac{e^{x/2}}{(\pi^{2} n x)^{1/4}}\cos \left[2
\sqrt{nx} - \pi/4\right] + O (n^{-3/4}),
\end{equation}
for $x>0$ ,
\begin{equation}\label{asympLag2}
L_{n}(x)= e^{x/2}J_{0}(2\sqrt{(n+1/2)x} + O (n^{-3/4}),
\end{equation}
for $a\leq x \leq b$, $a > 0$, and
\begin{equation}\label{asympLag3}
L_{n}(x)= 1 - nx + O \left((n x)^2\right),
\end{equation}
for $n x \rightarrow 0$. Therefore, we get, for some
$D_1(k),D_2(k)>0$,  the estimates :
\begin{eqnarray}\label{estimDelta3}
&&\left|\stackunder{n>n_{0}(V)}{\stackrel{+\infty }{\tsum
}}\left[\{L_{n}\left( \frac{\left|
h_{k}\right|^{2}}{2V}\right)-1\} e^{-\beta \left[ \left(
\varepsilon _{k}-\mu - \frac{g_{k}}{2V}\right)n \right]}
\left(\frac{\beta n^{2}}{2V}\right)\right]
\stackunder{n=0}{ \stackrel{+\infty }{\tsum }}e^{-\beta
\left[ \left( \varepsilon _{k}-\mu - \frac{g_{k}}{2V}\right)
n\right]}\right|
\nonumber\\
&&\leq D_1(k) \frac{\beta(
n_{0}(V))^{7/4}}{V^{3/4}}e^{-\beta(\varepsilon_{k}-\mu)
n_{0}(V)}f(\beta(\varepsilon _{k}-\mu)),
\end{eqnarray}
and
\begin{eqnarray}\label{estimDelta4}
&&\left|\stackunder{n>n_{0}(V)}{\stackrel{+\infty }{\tsum
}}\left[\{L_{n}\left( \frac{\left|
h_{k}\right|^{2}}{2V}\right)-1\} e^{-\beta \left[ \left(
\varepsilon _{k}-\mu - \frac{g_{k}}{2V}\right)n \right]}\right]
\stackunder{n=0}{ \stackrel{+\infty }{\tsum }}\left[e^{-\beta
\left[ \left( \varepsilon _{k}-\mu - \frac{g_{k}}{2V}\right)
n\right]}\left(\frac{\beta n^{2}}{2V}\right)\right]\right|
\nonumber\\
&&\leq D_2(k) \frac{\beta}{V^{3/4}}e^{-\beta(\varepsilon
_{k}-\mu)n_{0}(V)}\partial_{y}^{2}f(y=\beta(\varepsilon
_{k}-\mu)).
\end{eqnarray}
Since $\mu<0$, then for large $V$ the estimates
(\ref{estimDelta3}) and (\ref{estimDelta4}) are of the order
$O(e^{-\beta(\varepsilon _{k}-\mu) V^{(1-\delta)}})$ for
some $0 < \delta < 1$. Taking into account (\ref{estimDelta1}) and
(\ref{estimDelta2}) one concludes that $\Delta_{k}(V)$, and
consequently $(\Gamma_k-\gamma_k)$, have for large $V$ the order
$O(e^{-\beta\varepsilon _{k}}V^{-2})$. This implies that for any
$h\in\mathcal{D}$
\begin{equation} \label{limit}
\stackunder{\Lambda }{\lim } \stackunder{k\in \Lambda
^{*}\backslash \left\{ 0\right\} }{\prod } \Gamma_k
=\stackunder{\Lambda }{\lim } \stackunder{k\in \Lambda
^{*}\backslash \left\{ 0\right\} }{\prod } \gamma_k \left(1
+O(e^{-\beta\varepsilon
_{k}}V^{-2})\right)=\stackunder{\Lambda }{\lim }
\stackunder{k\in \Lambda ^{*}\backslash \left\{ 0\right\} }{\prod
} \gamma_k ,
\end{equation}
which,  by (\ref{newCCR23}), proves the assertion
(\ref{CCR23}).$\blacksquare $

\begin{remark}
\label{CCRthsup0} The conditions $0\leq g_{-}\leq g_{k}\leq g_{+}$
for $k\in \Lambda ^{*}\backslash \left\{ 0\right\} $ can be
relaxed. If $0\leq g_{k}=g_{k}\left( V\right) \leq \gamma
_{k}V^{\alpha _{k}}$ for $k\in \Lambda ^{*}\backslash \left\{
0\right\} ,$ with $\alpha _{k}\leq \alpha _{+}<1$ and $0\leq
\gamma _{k}\leq \gamma _{+}$, then (\ref{CCR23}) still holds. The
proof is obtained by following the same line of reasoning as in the 
proof of Theorem \ref{CCRth9}.
\end{remark}
\begin{remark}
\label{CCRthsup1}The first result of the Theorem \ref{CCRth9},
i.e. (\ref {CCR17}), is similar to the result for the PBG at
densities $\rho \geq \rho _{c}^{P}\left( \beta \right) $ (\ref{densite
critique du PBG}) in the {\it canonical} ensemble $\left( \beta
,\rho \right) $ (see \cite {ArakiWoods1,Cannon1}):
\[
\stackunder{\Lambda }{\lim }\left\langle
e^{\frac{i}{\sqrt{2V}}\left(
\overline{h_{0}}a_{0}+h_{0}a_{0}^{*}\right) }\right\rangle
_{T_{\Lambda
}}\left( \beta ,\rho \right) \equiv \stackunder{\Lambda }{\lim }\frac{\Tr_{%
\mathcal{H}_{B}^{\left( n\right) }}\left( \left\{ e^{\frac{i}{\sqrt{2V}}%
\left( \overline{h_{0}}a_{0}+h_{0}a_{0}^{*}\right) }e^{-\beta
T_{\Lambda }}\right\} ^{\left( n\right) }\right)
}{\Tr_{\mathcal{H}_{B}^{\left( n\right)
}}\left( \left\{ e^{-\beta T_{\Lambda }}\right\} ^{\left( n\right) }\right) }%
=J_{0}\left( \sqrt{2\rho _{0}^{P}\left( \beta ,\rho \right)
}\left| \widehat{h}_{0}\right| \right) ,
\]
where $n=\left[ V\rho \right] $ is the integer part of $V\rho $ and $%
\mathcal{H}_{B}^{\left( n\right) }$ is defined by(\ref{definition
de Hilbert space for n part}). Here
\[
\rho _{0}^{P}\left( \beta ,\rho \right) =\rho -\rho _{c}^{P}\left(
\beta
\right) =\stackunder{\Lambda }{\lim }\left\langle \frac{a_{0}^{*}a_{0}}{V}%
\right\rangle _{T_{\Lambda }}\left( \beta ,\rho \right) \equiv \stackunder{%
\Lambda }{\lim }\frac{1}{V}\frac{\Tr_{\mathcal{H}_{B}^{\left( n\right)
}}\left( \left\{ a_{0}^{*}a_{0} e^{-\beta T_{\Lambda
}}\right\} ^{\left( n\right) }\right)
}{\Tr_{\mathcal{H}_{B}^{\left( n\right) }}\left( \left\{ e^{-\beta
T_{\Lambda }}\right\} ^{\left( n\right) }\right) }
\]
is the \textit{canonical} density of conventional BE condensate
in the PBG. It should be stressed, however, that the limit
(\ref{CCR17}) is computed in the grand-canonical ensemble $\left(
\beta ,\mu \right) $ for the non-PBG (\ref{diagmodel1}).
\end{remark}

\begin{corollary}
\label{CCRth6}Let $0\leq g_{k}=g_{k}\left( V\right) \leq \gamma
_{k}V^{\alpha _{k}}$ for $k\in \Lambda ^{*}\backslash \left\{
0\right\} ,$ with $\alpha _{k}\leq \alpha _{+}<1$ and $0\leq
\gamma _{k}\leq \gamma _{+}$. Then for $ \varepsilon _{0}\in
\mathbb{R}^{1},$ $\mu <0$, and $h\in \mathcal{D}$, one has
\begin{eqnarray}
\mathbb{E}^{I}\left( \beta ,\mu ;h\right) \equiv \text{
}\stackunder{\Lambda }{\lim }\mathbb{E}_{\Lambda }^{I}\left( \beta
,\mu ;h\right) =J_{0}\left( \sqrt{ 2\rho _{0}^{I}\left( \beta ,\mu
\right) }\left| \widehat{h}_{0}\right| \right) \exp \left\{
-\frac{1}{4}\left\| h\right\| ^{2}-\frac{1}{2}A_{\beta ,\mu
}\left(
h,h\right) \right\} ,  \nonumber \\
&&  \label{CCR14bis}
\end{eqnarray}
where $\rho _{0}^{I}\left( \beta ,\mu \right) $ and $A_{\beta ,\mu
}\left( h_{1},h_{2}\right) $ are respectively the
non-conventional Bose condensate density (\ref{Generating0}) and
the positive {\it closable} sesquilinear form (\ref{CCR11bis}) with
domain $\mathcal{D}$.
\end{corollary}
\noindent \textit{Proof.} See Theorem \ref{CCRth9} and
Remark \ref{CCRthsup0}.

The last result is obtained for any \textit{fixed} chemical
potential $\mu <0$. It can be shown \cite{BruZagrebnov5} that for
\textit{finite} volume and any  density $\rho \geq 0$ there is a
one-to-one correspondence between $\rho$ and chemical potential $\mu _{\Lambda }^{I}\left(\beta, \rho \right)$, which is  solution of the equation
\begin{equation}
\rho _{\Lambda }^{I}\left( \beta ,\mu \right) \equiv \left\langle \frac{%
N_{\Lambda }}{V}\right\rangle _{H_{\Lambda }^{I}}\left( \beta
,\mu \right) =\rho .  \label{Generating4.0}
\end{equation}
{From} Corollary \ref{CCRth6} we have an explicit calculation of
the grand-canonical generating functional $\mathbb{E}_{\Lambda
}^{I}\left( \beta ,\mu ;h\right) $ in the thermodynamic limit.
However, for a \textit{fixed} total particle density $\rho \geq 0$
one has to evaluate the following thermodynamic limit
\begin{equation}
\widetilde{\mathbb{E}}^{I}\left( \beta ,\rho ;h\right) \equiv \text{ }%
\stackunder{\Lambda }{\lim }\mathbb{E}_{\Lambda }^{I}\left( \beta
,\mu
_{\Lambda }^{I}\left(\beta, \rho \right) ;h\right) ,\text{ }h\in \mathcal{D}\text{.
}  \label{CCRnew0}
\end{equation}
This is \textit{not} guaranteed to equal the limit (\ref{CCR14bis}) with
chemical
potential $\mu ^{I}\left(\beta, \rho \right) \equiv \stackunder{\Lambda }{\lim }%
\mu_{\Lambda }^{I}\left(\beta, \rho \right) $, since the map
$\rho \rightarrow \mu^{I}\left(\beta, \rho \right)$ fails to be injective:
\begin{equation}\label{Generating4bis}
\mu ^{I}\left(\beta, \rho \right) =\text{ }\stackunder{\Lambda }{\lim
}\mu _{\Lambda }^{I}\left(\beta, \rho \right) =\left\{
\begin{array}{c}
<0\text{ for }\rho <\rho _{c}^{I}\left( \beta \right) \\
=0\text{ for }\rho \geq \rho _{c}^{I}\left( \beta \right)
\end{array}
\right. ,
\end{equation}
because of conventional BE condensation for dimensions $
d > 2$. In this case the total particle density $\rho ^{I}\left( \beta
,\mu \right) $ (\ref{Generating01}) is saturated for $\mu =0$ \cite{BruZagrebnov5}:
there is a \textit{finite} critical density of particles $\rho _{c}^{I}\left( \theta
\right) $ , cf. (\ref {densite critique du PBG}),
(\ref{Generating1}). The question of equality of the limits 
(\ref{CCR14bis}) and (\ref{CCRnew0})
will be considered in the following paragraphs.

In fact it is linked to another question, which concerns the relation between
thermodynamic limit of the \textit{canonical} generating functional defined for
the total particle density $\rho \geq 0$ by
\begin{equation}
\mathbb{E}_{\Lambda ,can}^{I}\left( \beta ,\rho ;h\right) \equiv \frac{\Tr_{%
\mathcal{H}_{B}^{\left( n\right) }}\left( \left\{
W^{\mathcal{F}_{\Lambda }^{B}}\left( h\right) e^{-\beta
H_{\Lambda }^I}\right\} ^{\left( n\right) }\right)
}{\Tr_{\mathcal{H}_{B}^{\left( n\right) }}\left( \left\{ e^{-\beta
H_{\Lambda}^I}\right\} ^{\left( n\right) }\right) },\text{ }h\in \mathcal{D}%
_{\Lambda },\text{ }n=\left[ V\rho \right] ,
\label{CCR5canonique}
\end{equation}
and the \textit{grand-canonical} generating functional $\widetilde{\mathbb{E}}%
^{I}\left( \beta ,\rho ;h\right) $ defined by (\ref{CCRnew0}). In other words,
in the thermodynamic limit, for the same particle density $\rho $, the
\textit{canonical} ensemble may \textit{not} yield the same equilibrium
state as the \textit{grand-canonical} ensemble, i.e. one may have 
\[
\mathbb{E}_{can}^{I}\left( \beta ,\rho ;h\right) \equiv \stackunder{\Lambda }{%
\lim }\mathbb{E}_{\Lambda ,can}^{I}\left( \beta ,\rho ;h\right) \neq \widetilde{%
\mathbb{E}}^{I}\left( \beta ,\rho ;h\right) ,\text{ }h\in
\mathcal{D}\text{.}
\]

To answer to these questions we notice that
\begin{equation}
\mathbb{E}_{\Lambda }^{I}\left( \beta ,\mu _{\Lambda }^{I}\left(
\rho \right) ;h\right) =\frac{\stackunder{n=0}{\stackrel{+\infty
}{\sum }}e^{-\beta V\left[ -\mu _{\Lambda }^{I}\left( \rho
\right) \left( \frac{n}{V}\right) +f_{\Lambda }^{I}\left( \beta
,\frac{n}{V}\right) \right] }\mathbb{E}_{\Lambda
,can}^{I}\left( \beta ,\frac{n}{V};h\right) }{\stackunder{n=0}{\stackrel{%
+\infty }{\sum }}e^{-\beta V\left[ -\mu _{\Lambda }^{I}\left(
\rho \right) \left( \frac{n}{V}\right) +f_{\Lambda }^{I}\left(
\beta ,\frac{n}{V}\right) \right] }},\text{ }n=\left[ V\rho
\right] ,  \label{CCR44}
\end{equation}
where $f_{\Lambda }^{I}\left( \beta ,\rho \right) $ is the
free-energy density associated with the Hamiltonian $ H_{\Lambda
}^{I}$ (\ref{diagmodel1}):
\begin{equation}
f_{\Lambda }^{I}\left( \beta ,\rho \right) \equiv -\frac{1}{\beta V}\ln \Tr_{
\mathcal{H}_{B}^{\left( n\right) }}\left( \left\{ e^{-\beta
H_{\Lambda
}^{I}}\right\} ^{\left( n\right) }\right) ,\text{ }\rho \geq 0,\text{ }
n=\left[ V\rho \right] .  \label{diagmodelthese1}
\end{equation}
It is known from \cite{BruZagrebnov5} that :

\noindent (a)
$\{ f_{\Lambda }^{I}\left( \beta ,\rho \right)\}_{\Lambda\subset\mathbb{R}^d}$
is the family of \textit{strictly} convex functions of $\rho \geq 0$,
and this is also valid for
\begin{equation}
f^{I}\left( \beta ,\rho \right) \equiv \text{
}\stackunder{\Lambda }{\lim } f_{\Lambda }^{I}\left( \beta ,\rho
\right),  \label{Generating6}
\end{equation}
the free-energy density $f_{\Lambda }^{I}\left( \beta ,\rho
\right) $ ( \ref{diagmodelthese1}) in the thermodynamic limit,
but in the \textit{smaller} domain: $\rho < \rho
_{c}^{I}\left( \beta \right) $ ;

\noindent (b) the convergence (\ref{Generating6}) implies in this
domain (Griffiths lemma, see e.g. \cite{BruZagrebnov7}) the limit :
\begin{equation}
\partial_{\rho}f^{I}\left( \beta ,\rho \right)= \text{ }\stackunder{\Lambda }
{\lim }\partial_{\rho}f_{\Lambda }^{I}\left( \beta ,\rho \right)\equiv
\mu ^{I}_{can}\left( \beta,\rho \right) < 0 ,  \label{limDf}
\end{equation}
which coincides with (\ref{Generating4bis}):
\begin{equation}
\mu ^{I}_{can}\left( \beta,\rho \right)= \mu ^{I}\left(
\beta,\rho \right) , \label{limDf1}
\end{equation}
for $\rho < \rho _{c}^{I}\left( \beta \right)$ , where this
function is one-to-one ;

\noindent (c) the grand-canonical pressure,
$p^{I}\left(\beta,\mu\right) \equiv \stackunder{\Lambda }{\lim
}p^{I}_{\Lambda}\left(\beta,\mu\right)$, is the Legendre
transform:
\begin{equation} \label{Legendre}
p^{I}\left(\beta,\mu\right)=\stackunder{\rho\geq 0}{\sup }\left\{ \mu \rho
-f^{I}\left(
\beta ,\rho\right) \right\}= \left\{\mu \rho - f^{I}\left( \beta ,\rho
\right) \right\} \bigg|_{\rho
= \rho^{I}\left(\beta,\mu\right)},
\end{equation}
and
\begin{equation} \label{Legendre0}
f^{I}\left(\beta,\rho\right)=\stackunder{\mu\leq 0}{\sup }\left\{ \mu \rho
-p^{I}\left(
\beta ,\mu\right) \right\}= \left\{\mu \rho - p^{I}\left( \beta ,\mu
\right) \right\} \bigg|_{\mu
= \mu^{I}\left(\beta,\rho\right)},
\end{equation}
where
$\rho^{I}\left(\beta,\mu<0\right)<\rho^{I}_{c}\left(\beta\right)$
is the function inverse to the injection (\ref{limDf1}):
\[
\rho ^{I}\left( \beta ,\mu ^{I}\left(\beta, \rho
\right) \right) =\rho ;
\]
\noindent (d) for $\rho\geq\rho^{I}_{c}\left(\beta\right)$,
see (\ref{Generating2})
and (\ref{Generating4bis}), the limit (\ref{Generating6}) is equal to
\begin{equation}\label{Legendre1}
f^{I}\left(\beta,\rho\right)=\stackunder{\mu\leq 0}{\sup }\left\{ \mu \rho
-p^{I}\left(
\beta ,\mu\right) \right\}=  - p^{I}\left( \beta ,\mu =0
\right),
\end{equation}
i.e. the free-energy density is \textit{not} a strictly convex function in
this domain; respectively, $p^{I}\left(\beta,\mu=0\right)=
\stackunder{\rho\geq 0}{\sup }\left\{
-f^{I}\left(
\beta ,\rho\right) \right\}=  - f^{I}\left(\beta ,\rho\geq\rho^{I}_{c}\left(\beta\right)
\right)$, which means that the pressure and the free-energy density are always
related by the Legendre transform: \textit{weak equivalence} of ensembles.

By virtue of (a)-(c) we can now apply the Laplace \textit{large deviation
principle }\cite{Lewis1,LewisPfister1} to calculate the limit of
(\ref{CCR44}) in domain $\rho <\rho _{c}^{I}(\beta)$:
\begin{equation}
\widetilde{\mathbb{E}}^{I}\left( \beta ,\rho
;h\right) =\mathbb{E}^{I}\left( \beta ,\mu ^{I}\left(\beta,\rho
\right);h\right) =\mathbb{E}_{can}^{I}\left( \beta ,\rho ;h\right) .
\label{CCR45}
\end{equation}
Notice that for $d=1,2$ one has $\rho _{c}^{I}=+\infty$, see
(\ref{Generating1}). Therefore, Corollary \ref{CCRth6} together
with (\ref{CCR45}) imply the following result.

\begin{theorem}
\label{CCRth6bisbis} Let $\rho <\rho _{c}^{I}\left( \beta \right)
$. Then
\begin{eqnarray}
\widetilde{\mathbb{E}}^{I}\left( \beta ,\rho <\rho _{c}^{I}\left(
\beta \right) ;h\right) &=&J_{0}\left( \sqrt{2\rho _{0}^{I}\left(
\theta ,\mu ^{I}\left(\beta, \rho \right) \right) }\left|
\widehat{h}_{0}\right| \right) \exp \left\{ - \frac{1}{4}\left\| h\right\|
^{2}-\frac{1}{2}A_{\beta ,\mu ^{I}\left(\beta, \rho
\right) }\left( h,h\right) \right\} ,  \nonumber \\
&&  \label{CCR29bis}
\end{eqnarray}
for $d\geq 1$, $\varepsilon _{0}\in \mathbb{R}^{1}$, $0\leq
g_{k}\leq \gamma _{k}V^{\alpha _{k}}$ for $k\in \Lambda
^{*}\backslash \left\{ 0\right\} ,$ with $\alpha _{k}\leq \alpha
_{+}<1$ and $0\leq\gamma _{k}\leq \gamma _{+}$ and $h\in
\mathcal{D}$. Here $A_{\beta ,\mu }\left( h_{1},h_{2}\right) $
and $ \mu ^{I}\left(\beta, \rho <\rho _{c}^{I}\left( \beta
\right) \right) <0$ are defined respectively by (\ref{CCR11bis})
and (\ref{Generating4bis}).
\end{theorem}

Consequently, the equality (\ref{CCR45}) shows the \textit{strong equivalence%
} between the canonical ensemble $\left( \beta ,\rho \right) $
and the grand-canonical ensemble $\left( \beta ,\mu \right) $ for
$\rho <\rho _{c}^{I}\left( \beta \right) $ (i.e. $\mu <0$): in the
$H_{\Lambda }^{I}$ model (\ref{diagmodel1}) for a fixed total
particle density $\rho <\rho _{c}^{I}\left( \beta \right) $ the
Gibbs state in the grand-canonical ensemble coincides with the
one in the canonical ensemble.

However, contrary to the \textit{non-conventional} Bose
condensation (\ref{Generating0}) the \textit{conventional} BE condensation $\widetilde{\rho }%
_{0}^{I}\left( \beta ,\rho \right)> 0$ (\ref{Generating2})
violates this strong equivalence. Indeed, by virtue of (d),
see (\ref{Legendre1}),
the limiting measure (\ref{CCR44}) relating two generating
functionals is
\textit{not degenerate} in domain $\rho \geq
\rho _{c}^{I}\left( \beta \right)$.
Therefore, similar to the PBG
\cite{ArakiWoods1}-\cite{Lewis2}, the existence of
the critical density $\rho _{c}^{I}\left( \beta \right) $
implies that for $\rho > \rho _{c}^{I}\left( \beta \right) $ one has
\begin{equation}
\widetilde{\mathbb{E}}^{I}\left( \beta ,\rho ;h\right)  \neq \mathbb{E}_{can}^{I}\left(
\beta ,\rho ;h\right) . \label{CCRfinsection0}
\end{equation}
In the next Section \ref{Section 4} we show that, in contrast
to (\ref{CCR45}), for $\rho > \rho _{c}^{I}\left( \beta \right) $
one also gets
\begin{equation}\label{CCRfinsection}
\widetilde{\mathbb{E}}^{I}\left( \beta ,\rho ;h\right)  \neq
\mathbb{E}^{I}\left( \beta ,0 ;h\right).
\end{equation}

\Section{Gibbs states and conventional condensation of type III \label
{Section 4}}

Since $\rho _{c}^{I}\left( \beta \right)< +\infty$ (\ref
{Generating1}) only for $d>2$, we consider $\Lambda\subset \mathbb{R}^{d>2}$.
In the interest of simplicity
we restrict ourselves to a cubic box of the volume $V=\left|\Lambda\right|=L^{d}$,
and we put $g_{k}=g\geq0$, $k\in\Lambda^*$.
Notice that our reasoning in the proofs of Theorems \ref{CCRth9}
and \ref{CCRth6bisbis} used that $\mu < 0$, and that
$\mu_{\Lambda }^{I}\left(\beta, \rho < \rho _{c}^{I}\left( \beta
\right)  \right) < 0$, for large $V$. For $\rho \geq \rho
_{c}^{I}\left( \beta \right)$ it is not the case, see
\ref{AppendixCCR A}. This difference modifies essentially the
calculations of thermodynamic limit of the generating functional.

$1{^{\circ }}.$ Our first step is to refine, for $V \rightarrow
+\infty$, the asymptotics of the chemical potential $ \mu
_{\Lambda }^{I}\left(\beta, \rho \right) $, which is solution of
equation (\ref{Generating4.0}):
\[
\rho _{\Lambda }^{I}\left( \beta ,\mu _{\Lambda }^{I}\left(\beta,
\rho \right) \right) =\frac{1}{V}\stackunder{k\in \Lambda
^{*}\backslash \left\{ 0\right\} }{\sum }\left\langle
N_{k}\right\rangle _{H_{\Lambda }^{I}}\left( \beta ,\mu _{\Lambda
}^{I}\left(\beta, \rho \right) \right) +\left\langle \frac{
N_{0}}{V}\right\rangle _{H_{\Lambda }^{I}}\left( \beta ,\mu
_{\Lambda }^{I}\left(\beta, \rho \right) \right) =\rho .
\]
For a \textit{strictly} positive $g>0$ this is done in \ref{AppendixCCR A},
see Theorem \ref{theoremAsymp}:
\begin{equation}
\mu _{\Lambda }^{I}\left(\beta, \rho \right) =\left( \frac{\rho
-\rho _{c}^{I}\left( \beta \right) }{CV}\right) ^{2/(d+2)}+O\left(
\frac{1}{V}\right) ,  \label{CCRnew9}
\end{equation}
where the constant $C= \left(2m/\hbar ^{2}\right)^{d/2}/\left[g\,\, 2^{d-2}\pi^{d/2}d(d+2)\Gamma(d/2)
\right]>0$, and $\Gamma(z)$ is the Euler gamma function. Hence, the
chemical potential $\mu_{\Lambda }^{I}\left(\beta, \rho \geq
\rho _{c}^{I}\left( \beta \right)  \right)$, is \textit{non-negative}
for large $V$.
By virtue of (\ref{CCR16}) and (\ref{CCR25}) this observation motivates
to represent the generating functional (\ref{CCR22}) in the form :
\begin{eqnarray}
\mathbb{E}_{\Lambda }^{I}\left( \beta ,\mu _{\Lambda
}^{I}\left(\beta, \rho \right) ;h\right) &=&\left\langle
e^{\frac{i}{\sqrt{2V}}\left( \overline{h_{0}}
a_{0}+h_{0}a_{0}^{*}\right) }\right\rangle _{H_{0}^{I}}\left(
\beta ,\mu
_{\Lambda }^{I}\left(\beta, \rho \right) \right)  \nonumber \\
&&\times \stackunder{k\in D_{+}^{\left( \Lambda \right) }}{\prod }
\left\langle e^{\frac{i}{\sqrt{2V}}\left( \overline{h_{k}}
a_{k}+h_{k}a_{k}^{*}\right) }\right\rangle _{H_{k}^{I}}\left(
\beta ,\mu
_{\Lambda }^{I}\left(\beta, \rho \right) \right)  \nonumber \\
&&\times \stackunder{k\in D_{-}^{\left( \Lambda \right) }}{\prod }%
\left\langle e^{\frac{i}{\sqrt{2V}}\left( \overline{h_{k}}%
a_{k}+h_{k}a_{k}^{*}\right) }\right\rangle _{H_{k}^{I}}\left(
\beta ,\mu _{\Lambda }^{I}\left(\beta, \rho \right) \right) ,
\label{CCRnew6}
\end{eqnarray}
with
\begin{equation}
\begin{array}{l}
D_{-}^{\left( \Lambda \right) }\equiv \left\{ k\in \Lambda
^{*}\backslash
\left\{ 0\right\} :\varepsilon _{k}-
\mu _{\Lambda }^{I}\left(\beta, \rho \right) -
\frac{g_{k}}{2V}< 0\right\} , \\
D_{+}^{\left( \Lambda \right) }\equiv \left\{ k\in \Lambda
^{*}\backslash \left\{ 0\right\} :\varepsilon _{k}- \mu _{\Lambda
}^{I}\left(\beta, \rho \right) - \frac{g_{k}}{2V}\geq 0\right\} .
\end{array}
\label{CCRnew1bisbisbis}
\end{equation}
\begin{remark}
\label{CCRremarkplus} As it is shown in \ref{AppendixCCR A}, for
$\rho >\rho _{c}^{I}\left( \beta \right) $ and  $g_k \geq g > 0$
the non-extensive condensation $\widetilde{\rho }_{0}^{I}\left(
\beta ,\rho \right) $ (\ref{Generating2}) is concentrated on the set
$D_{-}^{\left( \Lambda \right) }$:
\begin{equation}
\widetilde{\rho }_{0}^{I}\left( \beta ,\rho \right) =\text{ }\stackunder{%
\Lambda }{\lim }\frac{1}{V}\stackunder{k\in D_{-}^{\left( \Lambda \right) }}{%
\sum }\left\langle N_{k}\right\rangle _{H_{\Lambda }^{I}}\left(
\beta ,\mu _{\Lambda }^{I}\left(\beta, \rho \right) \right) =\rho
-\rho _{c}^{I}\left( \beta \right) >0,  \label{Generating2bis}
\end{equation}
see Lemma \ref{theoremCCR1} in \ref{AppendixCCR A}.
\end{remark}

$2{^{\circ }}.$ Since in the proof of Theorem \ref{CCRth9}$\left(
i\right)$ the sign of $\mu$ is irrelevant, the same line of
reasoning gives that for $\mu =\mu _{\Lambda }^{I}\left(\beta,
\rho\geq \rho _{c}^{I}\left( \beta
\right) \right)$, (\ref{CCRnew9}), and  for any $ h\in \mathcal{D}$ ,
$\varepsilon
_{0}\in \mathbb{R}^{1}$:
\begin{equation}
\stackunder{\Lambda }{\lim }\left\langle
e^{\frac{i}{\sqrt{2V}}\left(
\overline{h_{0}}a_{0}+h_{0}a_{0}^{*}\right) }\right\rangle
_{H_{0}^{I}}\left( \beta ,\mu _{\Lambda }^{I}\left(\beta, \rho \right)
\right) =J_{0}\left( \sqrt{2\rho _{0}^{I}\left( \beta ,0\right)
}\left| \widehat{h}_{0}\right| \right) .  \label{CCRplus0}
\end{equation}
Here the non-conventional Bose-condensate density $\rho
_{0}^{I}\left( \theta ,0\right) \ $ is defined for $\mu =0$ by
(\ref{Generating0}).

$3{^{\circ }}.$ Now, using the asymptotics (\ref{CCRnew9}) we can
evaluate the thermodynamic limit of
\begin{equation}
\stackunder{k\in D_{+}^{\left( \Lambda \right) }}{\prod }\left\langle e^{%
\frac{i}{\sqrt{2V}}\left(
\overline{h_{k}}a_{k}+h_{k}a_{k}^{*}\right) }\right\rangle
_{H_{k}^{I}}\left( \beta ,\mu _{\Lambda }^{I}\left(\beta, \rho \right)
\right) ,  \label{CCRplus}
\end{equation}
see (\ref{CCRnew6}). Indeed, by inspection of the line of reasoning
(\ref{CCR25})-(\ref{estimDelta4}) we find that it is only the
inequality $\varepsilon _{k}- \mu _{\Lambda }^{I}\left(\beta,
\rho \right) - \frac{g_{k}}{2V}>0$ which one needs that the limit
(\ref{limit}) be valid. Therefore, taking into account
(\ref{CCRnew9}) we get
\begin{eqnarray}
\stackunder{\Lambda }{\lim }\stackunder{k\in D_{+}^{\left( \Lambda \right) }%
}{\prod }\left\langle e^{\frac{i}{\sqrt{2V}}\left( \overline{h_{k}}%
a_{k}+h_{k}a_{k}^{*}\right) }\right\rangle _{H_{k}^{I}}
\left( \beta ,\mu _{\Lambda }^{I}\left(\beta, \rho \right)
\right) &=&\exp \left\{ -
\frac{1}{4}\left\| h\right\| ^{2}-\frac{1}{2}A_{\beta ,0}\left(
h,h\right)
\right\} ,  \nonumber \\
&&  \label{CCRnew7}
\end{eqnarray}
for $h\in \mathcal{D}$ and $\rho \geq \rho_{c}^{I}\left( \theta
\right) $, with the sesquilinear form $A_{\beta ,\,\mu =0}\left(
h_{1},h_{2}\right) $ defined by (\ref{CCR11bis}).

$4{^{\circ }}.$ Finally, using the asymptotics (\ref{CCRnew9}) we have to
compute the thermodynamic limit of the
last factor in (\ref{CCRnew6}):
\begin{equation}
\stackunder{\Lambda }{\lim }\stackunder{k\in D_{-}^{\left( \Lambda \right) }}{\prod }\left\langle e^{%
\frac{i}{\sqrt{2V}}\left(
\overline{h_{k}}a_{k}+h_{k}a_{k}^{*}\right) }\right\rangle
_{H_{k}^{I}}\left( \beta ,\mu _{\Lambda }^{I}\left(\beta, \rho \right)
\right) .  \label{CCRplusplus}
\end{equation}

Because of \textit{non-negative} $ \mu _{\Lambda
}^{I}\left(\beta, \rho \right)$, and of the
\textit{non-extensive} BE condensation $\widetilde{\rho
}_{0}^{I}\left( \theta ,\rho \right) $ (\ref{Generating2bis}),
that spreads over the modes $k\in D_{-}^{\left( \Lambda \right) }$
( Remark \ref{CCRremarkplus} ), this calculation is a more subtle
matter than the thermodynamic limit of (\ref{CCRplus}). In
particular, the exact knowledge of the asymptotics
(\ref{CCRnew9}) of $\mu _{\Lambda }^{I}\left(\beta, \rho \right)
$ becomes essential to find the limit of (\ref{CCRplusplus}).

By (\ref{CCR25}) we get:
\begin{equation} \label{CCRnew4}
\Gamma_k = \left\langle e^{\frac{i}{\sqrt{2V}}\left(
\overline{h_{k}} a_{k}+h_{k}a_{k}^{*}\right) }\right\rangle
_{H_{k}^{I}} = e^{-\frac{1}{4V}\left|
h_{k}\right|^{2}}\stackunder{n=0}{\stackrel{ +\infty }{\sum }}\nu
_{\Lambda ,k}\left( \beta ,\rho \,; n\right)
L_{n}\left(\frac{\left| h_{k}\right|^{2}}{2V} \right),
\end{equation}
where
\begin{equation} \label{measure0}
\nu _{\Lambda ,k}\left( \beta,\rho\, ;n \right) \equiv\frac{e^{-\beta \left[
\left(
\varepsilon _{k}-\mu _{\Lambda }^{I}\left(\beta, \rho \right) -\frac{g_{k}}{2V}
\right) n+\frac{g_{k}}{2V}n^{2}\right] }}{\stackunder{n=0}{
\stackrel{+\infty }{\sum }} e^{-\beta \left[ \left( \varepsilon
_{k}-\mu
_{\Lambda }^{I}\left(\beta, \rho \right) -\frac{g_{k}}{2V}\right) n+\frac{g_{k}}{2V}
n^{2}\right]}}
\end{equation}
is the family of probability measures $\{ \nu _{\Lambda ,k}\left( \beta,\rho\,
;n \right) \}_ {\Lambda\subset\mathbb{R}^{d} ,\, k\in\Lambda^{*}}$ , cf. (\ref{distr})
and (\ref{measure}).
Consequently
\begin{equation} \label{sum}
\ln \stackunder{k\in D_{-}^{\left( \Lambda \right) }}{\prod
}\Gamma_k = \left(-\frac{1}{4V}\right)\stackunder{k\in D_{-}^{\left( \Lambda
\right) }}{\sum }\left| h_{k}\right| ^{2} + \stackunder{k\in
D_{-}^{\left( \Lambda \right) }}{\sum
}\ln\stackunder{n=0}{\stackrel{ +\infty }{\sum }}\nu _{\Lambda
,k}\left( \beta ,\rho \,; n\right) L_{n}\left(\frac{\left|
h_{k}\right| ^{2}}{2V} \right).
\end{equation}
Since $\mu _{\Lambda }^{I}\left(\beta, \rho \geq\rho_{c}^{I}(\beta)\right)\rightarrow 0
$, the thermodynamic limit of the first term in (\ref{sum}) is
\begin{equation} \label{sum1}
\stackunder{\Lambda }{\lim }\left(-\frac{1}{4V}\right)\stackunder{k\in
D_{-}^{\left( \Lambda \right) }}{\sum }\left| h_{k}\right| ^{2}=
\stackunder{\delta\rightarrow 0}{\lim
}\left(-\frac{1}{4}\right)\frac{1}{(2\pi)^d}\stackunder{\{k:
\varepsilon_{k}\leq \delta\} }{\int } d^{d}k \left| \widehat{h}_{k}
\right| ^{2}=0 ,
\end{equation}
for any $h\in\mathcal{D}$. By virtue of the asymptotics
(\ref{CCRnew9}) and by definition of domain $D_{-}^{\left(
\Lambda \right)}$ we get for the limit of the second term :
\begin{eqnarray}\label{sum2}
&&\stackunder{\Lambda }{\lim }\stackunder{k\in D_{-}^{\left(
\Lambda \right) }}{\sum }\ln\stackunder{n=0}{\stackrel{ +\infty
}{\sum }}\nu _{\Lambda ,k}\left( \beta ,\rho \,; n\right)
L_{n}\left(\frac{\left| h_{k}\right| ^{2}}{2V} \right)=
\nonumber\\
&&\stackunder{\Lambda }{\lim }\stackunder{s\in \mathcal{S}_B}{\sum
}\ln\stackunder{n=0}{\stackrel{ +\infty }{\sum }}\nu _{\Lambda ,
k(s) }\left( V^{1-2\gamma}\beta ,\rho \,; n/V^{1-\gamma}\right)
L_{n}\left(\frac{\left| h_{k(s)/V^{\gamma/2}}\right| ^{2}}{2V}
\right).
\end{eqnarray}
Here $\gamma=2/(d+2)$, the set :
\begin{equation}\label{S}
\mathcal{S}_B
\equiv\left\{s=\left\{s_{\alpha}\right\}_{\alpha=1}^{d}\in
\mathbb{Z}^{d}\backslash\left\{0\right\}: \frac{\hbar
^{2}}{2m}(2\pi)^{2} \stackunder{\alpha=1}{ \stackrel{d}{\sum }}
({s_{\alpha}}/{V^{1/d-\gamma/2}})^{2}\leq B\right\},
\end{equation}
and we put $k(s)\equiv 2\pi s/V^{1/d-\gamma/2}$. Since for $d>2$ one has
$1-2\gamma >0$, the family of the \textit{scaled} probability measures:
\begin{equation}\label{measureScaled}
\{
\nu _{\Lambda ,k(s)}\left(
V^{1-2\gamma}\beta,\rho\, ;n/V^{1-\gamma} \right) \}_
{\Lambda\subset\mathbb{R}^{d} ,\, s\in\mathcal{S_B}},
\end{equation}
see (\ref{measure0}), verifies the Laplace \textit{large deviation
principle} \cite{Lewis1,LewisPfister1} with the support at
the point $\stackunder{\Lambda }{\lim
}\overline{n}(V)/V^{1-\gamma} = (B-\varepsilon_k(s))/g$ , cf.
Theorem \ref{theoremCCR1}. This remark, together with the asymptotics
(\ref{asympLag3}) of the Laguerre polynomials and the continuity
of $h\in\mathcal{D}$, gives :
\begin{eqnarray}\label{limsum2}
&&\stackunder{\Lambda }{\lim }\stackunder{s\in \mathcal{S}_B}{\sum
}\ln\stackunder{n=0}{\stackrel{ +\infty }{\sum }}\nu _{\Lambda ,
k(s) }\left( V^{1-2\gamma}\beta ,\rho \,; n/V^{1-\gamma}\right)
L_{n}\left(\frac{\left| h_{k(s)/V^{\gamma/2}}\right| ^{2}}{2V}
\right) =\nonumber\\
&&\stackunder{\Lambda }{\lim }\stackunder{s\in \mathcal{S}_B}{\sum
}\ln
L_{\overline{n}(V) = V^{1-\gamma}(B-\varepsilon_k(s))/g}\left(\frac{\left| h_{k(s)/V^{\gamma/2}}\right| ^{2}}{2V}
\right) =
\nonumber\\
&&\stackunder{\Lambda }{-\lim }\stackunder{s\in
\mathcal{S}_B}{\sum }  V^{1-\gamma}
\frac{(B-\varepsilon_{k(s)})}{g} \frac{\left|
h_{k(s)/V^{\gamma/2}}\right| ^{2}}{2V} = -\left|
\widehat{h}_{0}\right| ^{2} \frac{1}{\left( 2\pi \right)
^{d}}\stackunder{\{k:\, \varepsilon_{k} \leq B\}}{\int }d^{d}k
\frac{B-\varepsilon_{k}}{g}=\nonumber\\
&&-\frac{1}{2}\left| \widehat{h}_{0}\right| ^{2}(\rho -\rho
_{c}^{I}\left(\beta \right)).
\end{eqnarray}
Here we used that $\gamma=1-d\gamma/2$ to obtain in the limit the integral,
and (\ref{B}) to get the last equality.
Taking (\ref{sum1}) and (\ref{limsum2}) into account we finally get
for (\ref{CCRplusplus}):
\begin{eqnarray}
&&\stackunder{\Lambda }{\lim } \stackunder{k\in
D_{-}^{\left( \Lambda
\right) }}{\prod }\left\langle e^{\frac{i}{\sqrt{2V}}\left( \overline{h_{k}}%
a_{k}+h_{k}a_{k}^{*}\right) }\right\rangle _{H_{k}^{I}}\left( \beta ,\mu _{\Lambda }^{I}\left(\beta, \rho \right)
\right)
=\exp \left\{ -\frac{1}{2}\left| \widehat{h}_{0}\right| ^{2}\left( \rho
-\rho _{c}^{I}\left( \beta
\right) \right)\right\}
\nonumber \\
&&=\exp \left\{ -\frac{1}{2}\left| \widehat{h}_{0}\right|
^{2}\widetilde{\rho }_{0}^{I}\left( \beta ,\rho \right) \right\} ,
\label{CCRsupsup3}
\end{eqnarray}
where $\widetilde{\rho }_{0}^{I}\left( \beta ,\rho \right)$ is
density of the \textit{type III} BE condensation, see
(\ref{Generating2}).

The results of $1{^{\circ}}- 4{^{\circ }}$  can be summed up as follows:
\begin{theorem}
\label{CCRIII} Let $\rho \geq\rho _{c}^{I}\left( \beta \right) $.
Then
\begin{eqnarray}
\widetilde{\mathbb{E}}^{I}\left( \beta ,\rho ;h\right) &\equiv
&\text{ } \stackunder{\Lambda }{\lim }\mathbb{E}_{\Lambda
}^{I}\left( \beta ,\mu
_{\Lambda }^{I}\left(\beta, \rho \right) ;h\right)  \nonumber \\
&=&J_{0}\left( \sqrt{2\rho _{0}^{I}\left( \beta ,0\right) }\left|
\widehat{h}_{0}\right| \right) \exp \left\{ -\frac{1}{4}\left\| h\right\| ^{2}-\frac{1%
}{2}A_{\beta ,0}\left( h,h\right) \right\}  \nonumber \\
&&\times \exp \left\{ -\frac{1}{2}\left| \widehat{h}_{0}\right| ^{2}\left(
\rho -\rho _{c}^{I}\left( \beta \right) \right) \right\},
\label{CCRnew10}
\end{eqnarray}
for $d>2$, $\varepsilon _{0}\in \mathbb{R}^{1}$, $g_{k}=g >0$ and
$h\in \mathcal{D}$. Here $A_{\beta ,\mu }\left(
h_{1},h_{2}\right) $ and $ \mu_{\Lambda } ^{I}\left(\beta, \rho
\geq\rho _{c}^{I}\left( \beta \right) \right) \geq 0$ are defined
respectively by (\ref{CCR11bis}) and (\ref{CCRnew9}).
\end{theorem}
\begin{corollary}
\label{corrCCRIII}
Comparing (\ref{CCR14bis}) and (\ref{CCRnew10})
one gets that the presence
of the \textit{conventional} BE condensation
$\widetilde{\rho }_{0}^{I}\left( \beta ,\rho \right)>0$, see
(\ref{Generating2}), (\ref{Generating2bis}), implies :
\begin{equation} \label{cond}
\widetilde{\mathbb{E}}^{I}\left( \beta ,\rho ;h\right)=
\mathbb{E}^{I}\left( \beta ,0;h\right)
\exp \left\{ -\frac{1}{2}\left| \widehat{h}_{0}\right|
^{2}\widetilde{\rho }_{0}^{I}\left( \beta ,\rho \right) \right\},
\end{equation}
cf. (\ref{CCRfinsection}).
This means that at the point $\mu = \stackunder{\Lambda }
{\lim }\mu_{\Lambda } ^{I}\left(\beta,
\rho\geq\rho _{c}^{I}\left( \beta \right) \right)=0$ the
grand-canonical equilibrium state is not unique.
There is a family of states
enumerated by the BE condensate density
$\widetilde{\rho }_{0}^{I}\left( \beta ,\rho \right)=
\rho -\rho _{c}^{I}\left( \beta \right)$.
\end{corollary}

This effect is well-known in the PBG \cite {Cannon1}-\cite{Lewis2}:
\begin{proposition}
\label{prop1}
For isotropic dilation of a rectangular container 
$\Lambda\subset\mathbb{R}^{d>2}$ the
grand-canonical generating functional
\begin{eqnarray} \label{PBGprop1}
&&\widetilde{\mathbb{E}}^{P}\left( \beta ,\rho ;h\right) \equiv
\stackunder{\Lambda }{\lim }\frac{\Tr_{\mathcal{F}_{\Lambda }^{B}}\left(\stackunder{k\in
\Lambda^{*}}{\prod } e^{
\frac{i}{\sqrt{2V}}\left(
\overline{h_{k}}a_{k}+h_{k}a_{k}^{*}\right) }e^{-\beta \left(
T_{\Lambda }-\mu _{\Lambda }^{P}\left(\beta, \rho \right) N_{\Lambda
}\right) }\right) }{\Tr_{\mathcal{F}_{\Lambda }^{B}}\left(
e^{-\beta \left( T_{\Lambda }-\mu _{\Lambda }^{P}\left(\beta, \rho
\right)
N_{\Lambda }\right) }\right) }= \nonumber \\
&&\stackunder{\Lambda }{\lim }\exp \left\{ -\frac{1}{4}\left\| h\right\| ^{2}-\frac{1%
}{2}A_{\beta ,\mu_{\Lambda }^{P}\left(\beta, \rho \right)}\left( h,h\right) \right\}\stackunder{\Lambda }{\lim }\left\langle
e^{\frac{i}{\sqrt{2V}}\left(
\overline{h_{0}}a_{0}+h_{0}a_{0}^{*}\right) }\right\rangle
_{T_{\Lambda
}}\left( \beta ,\mu _{\Lambda }^{P}\left(\beta, \rho \right) \right)
= \nonumber \\
&&\exp \left\{ -\frac{1}{4}\left\| h\right\| ^{2}-\frac{1
}{2}A_{\beta ,\mu ^{P}\left(\beta, \rho \right)}\left( h,h\right) \right\}
\exp \left\{ -\frac{1}{2}\left| \widehat{h}_{0}\right| ^{2}\rho
_{0}^{P}\left( \beta ,\rho \right) \right\}.
\end{eqnarray}
Here $\mu_{\Lambda }^{P}\left(\beta, \rho \right)<0$ is solution of
the equation
\[
\rho _{\Lambda }^{P}\left( \beta ,\mu \right) \equiv\left\langle
\frac{N_{\Lambda
}}{V}\right\rangle _{T_{\Lambda }}\left( \beta ,\mu \right) =
\frac{1}{V}\frac{\Tr_{
\mathcal{F}_{\Lambda }^{B}}\left( N_{\Lambda }e^{-\beta
\left( T_{\Lambda }-\mu N_{\Lambda }\right) }\right)
}{\Tr_{\mathcal{F}_{\Lambda
}^{B}}\left( e^{-\beta \left( T_{\Lambda }-\mu N_{\Lambda }\right) }\right) }%
=\rho.
\]
Similar to (\ref{Generating4bis}) and (\ref {Generating2bis}), one has:
\[
\mu ^{P}\left(\beta, \rho \right) =\text{ }\stackunder{\Lambda }{\lim
}\mu _{\Lambda }^{P}\left(\beta, \rho \right) =\left\{
\begin{array}{c}
<0\text{ for }\rho <\rho _{c}^{P}\left( \beta \right) \\
=0\text{ for }\rho \geq \rho _{c}^{P}\left( \beta \right)
\end{array}
\right. ,
\]
and the type I conventional BE condensation in the single mode $k=0$:
\begin{equation}\label{CCRsupsup5}
\widetilde{\rho }_{0}^{P}\left( \beta ,\rho \right)=
\stackunder{\Lambda }{\lim}\left\langle
\frac{a_{0}^{*}a_{0}}{V}\right\rangle _{T_{\Lambda
}}\left( \beta ,\mu _{\Lambda }^{P}\left(\beta, \rho \right) \right)
=\left\{
\begin{array}{c}
=\,0 \,\,\,\,\,\,\,\,\,\,\,\,\,\,\,\,\,\,\,\,\,\,\,\,\text{ for }\rho \leq\rho _{c}^{P}\left( \beta \right) \\
=\rho -\rho
_{c}^{P}\left( \beta \right)\text{ for }
\rho > \rho _{c}^{P}\left( \beta \right)
\end{array}
\right. .
\end{equation}
\end{proposition}
Hence, similar to (\ref{cond}), one has
\[
\widetilde{\mathbb{E}}^{P}\left( \beta ,\rho>\rho _{c}^{P}\left( \beta \right) ;h\right)=
{\mathbb{E}}^{P}\left( \beta ,0 ;h\right)\exp \left\{ -\frac{1}{2}\left| \widehat{h}_{0}\right| ^{2}\widetilde{\rho
}_{0}^{P}\left( \beta ,\rho \right) \right\}.
\]

Moreover, the remarks (a)-(d) are valid for PBG.
For $\rho\leq\rho _{c}^{P}\left( \beta \right)$ there is a
\textit{strong equivalence} of ensembles expressed by :
\begin{equation}\label{CCR45can}
\widetilde{\mathbb{E}}^{P}\left( \beta ,\rho
;h\right) =\mathbb{E}^{P}\left( \beta ,\mu ^{P}\left(\beta,\rho
\right);h\right) =\mathbb{E}_{can}^{P}\left( \beta ,\rho ;h\right),
\end{equation}
cf.(\ref{CCR45}). Whereas for $\rho>\rho _{c}^{P}\left( \beta \right)$,
the functional $\mathbb{E}_{can}^{P}\left( \beta ,\rho; h\right)$ and the
\textit{non-degenerate} measure in (\ref{CCR44}), relating the canonical
and the grand-canonical generating functionals,
can be calculated explicitly  \cite{Cannon1}-\cite{Lewis2}.
This gives :
\begin{equation}\label{Kac1}
\widetilde{\mathbb{E}}^{P}\left( \beta ,\rho ;h\right)=
\stackunder{\mathbb{R}^1}{\int } K_{\rho,\rho _{c}^{P}\left( \beta \right)}(dx)
\,\mathbb{E}_{can}^{P}\left( \beta ,x ;h\right).
\end{equation}
Here
\begin{equation}\label{Kac2}
\mathbb{E}_{can}^{P}\left( \beta, x >\rho _{c}^{P}\left( \beta \right);
h\right)=\exp \left\{ -\frac{1}{4}\left\| h\right\| ^{2}-\frac{1
}{2}A_{\beta ,0}\left( h,h\right) \right\}J_{0}\left( \sqrt{2(x-\rho _{c}^{P}
(\beta ))}\left|
\widehat{h}_{0}\right| \right),
\end{equation}
and
\begin{equation}\label{Kac}
K_{\rho,\rho _{c}^{P}\left( \beta \right)}(dx)=\left\{
\begin{array}{ll}
0&\text{for }x \leq\rho _{c}^{P}\left( \beta \right) \\
\{\rho-\rho _{c}^{P}\left( \beta \right)\}^{-1}\exp \left\{-\displaystyle\frac{x-\rho _{c}^{P}\left( \beta \right)}{\rho-\rho _{c}^{P}\left( \beta \right))}\right\} \,\,dx
&\text{for }x > \rho _{c}^{P}\left( \beta \right)
\end{array}
\right. ,
\end{equation}
is known as the \textit{Kac measure} for $\rho>\rho
_{c}^{P}\left( \beta \right)$. Since  by (a)-(c) the limiting
measure in (\ref{CCR44}) is \textit{degenerate} :
\begin{equation}\label{Kac0}
K_{\rho,\rho _{c}^{P}\left( \beta \right)}(dx) = \delta (x-\rho)\,dx ,
\end{equation}
for $\rho\leq\rho _{c}^{P}\left( \beta \right)$,
cf.(\ref{CCR45can}), the representation (\ref{Kac1}) is valid for
any $\rho \geq 0$.

\Section{Concluding remarks\label{Section 5}}

\textbf{5.1} To answer the question about the r\^{o}le of the
\textit{type }III BE condensation in determining the
generating functional (\ref{cond}), let us consider instead of
(\ref{diagmodel1}) a truncated model with the Hamiltonian
$H_{\Lambda}^{0}$.

Since Theorem \ref{CCRth9} is valid for
$g_{+} =0$ (or $\gamma_{+} =0$), the generating functional 
for the model $H_{\Lambda}^{0}$ has for $\mu<0$
the same form:
\begin{eqnarray} \label{trunc1}
\mathbb{E}^{0}\left( \beta ,\mu; h\right)
&=& \stackunder{\Lambda }{\lim}\stackunder{k\in
\Lambda ^{*}}{\prod }\left\langle e^{\frac{i}{
\sqrt{2V}}\left( \overline{h_{k}}a_{k}+h_{k}a_{k}^{*}\right)
}\right\rangle _{H_{\Lambda}^{0}}(\beta,\mu) \nonumber \\
&=&
J_{0}\left( \sqrt{ 2\rho _{0}^{0}\left( \beta ,\mu
\right) }\left| \widehat{h}_{0}\right| \right) \exp \left\{
-\frac{1}{4}\left\| h\right\| ^{2}-\frac{1}{2}A_{\beta ,\mu
}\left(h,h\right) \right\},
\end{eqnarray}
as for $H_{\Lambda}^{I}$, cf.(\ref{CCR14bis}). Here
$\rho _{0}^{0}\left( \beta ,\mu \right) = \rho _{0}^{I}\left( \beta ,\mu
\right)$, see (\ref{Generating0}). Again, in this domain we have
the \textit{strong equivalence} of ensembles (\ref{CCR45}):
\begin{equation}\label{trunc2}
\widetilde{\mathbb{E}}^{0}\left( \beta ,\rho
;h\right) =\mathbb{E}^{0}\left( \beta ,\mu ^{0}\left(\beta,\rho
\right);h\right) =\mathbb{E}_{can}^{0}\left( \beta ,\rho ;h\right),
\end{equation}
where $\mu ^{0}\left(\beta,\rho\right)=\mu ^{I}\left(\beta,\rho
\right)$ and $\rho _{c}^{0}\left( \beta \right)=
\rho _{c}^{I}\left( \beta \right)  $, see (\ref{Generating4bis}).

On the other hand, for $\rho>\rho _{c}^{0}\left( \beta \right)$
the model $H_{\Lambda}^{0}$ manifests (instead of the \textit{type} III) 
the BE condensation of the \textit{type} I, see \cite{BruZagrebnov5}.
More precisely, for dilatation of a \textit{cubic} $\Lambda$ there is the BE
condensation in $2d$ modes :
\[
\mathcal{K}_{2d} \equiv \{k\in\Lambda^* :
\{k_{i}^{\alpha}\}_{i=1}^{d}=(0,0,...,0,k_{\alpha}=\pm2\pi/V^{1/d},0,...,0),
\alpha=1,2,...,d \},
\]
such that
\begin{equation}\label{trunc3}
\stackunder{\Lambda }{\lim}\frac{1}{V}
\left\langle N_{k\in\mathcal{K}_{2d}}\right\rangle _{H_{\Lambda}^{0}}
(\beta,\mu_{\Lambda}^{0}(\beta,\rho>\rho _{c}^{0}\left( \beta \right)))
=\frac{1}{2d}(\rho-\rho _{c}^{0}\left( \beta \right)),
\end{equation}
and
\begin{equation}\label{trunc4}
\stackunder{\Lambda }{\lim}\frac{1}{V}
\left\langle N_{k\in\Lambda^{*}\setminus\{\mathcal{K}_{2d}
\cup\{0\}\}}\right\rangle _{H_{\Lambda}^{0}}
(\beta,\mu_{\Lambda}^{0}(\beta,\rho>\rho _{c}^{0}\left( \beta \right)))
=0 .
\end{equation}
It corresponds to the asymptotics :
\begin{equation}\label{trunc5}
\mu_{\Lambda}^{0}(\beta,\rho>\rho _{c}^{0}
\left( \beta \right))=\varepsilon_{k_{\alpha}}
- \frac{2d}{V\beta(\rho-\rho _{c}^{0}\left( \beta \right))}
+ O(\frac{1}{V})
\end{equation}
of the solution of the equation
\begin{equation}\label{trunc6}
\rho=\frac{1}{V}\stackunder{k\in\Lambda^{*}}
{\sum}\left\langle N_{k}\right\rangle _{H_{\Lambda}^{0}}
(\beta,\mu_{\Lambda}^{0}(\beta,\rho)).
\end{equation}
Since $\varepsilon_{k}-\mu_{\Lambda}^{0}(\beta,\rho)>0$ for $k\neq0$,
following the same line of reasoning as in calculation of (\ref{CCRnew6})
we get for :
\begin{eqnarray}\label{trunc7}
\mathbb{E}_{\Lambda }^{0}\left( \beta ,\mu _{\Lambda
}^{0}\left(\beta, \rho \right) ;h\right) &=&\left\langle
e^{\frac{i}{\sqrt{2V}}\left( \overline{h_{0}}
a_{0}+h_{0}a_{0}^{*}\right) }\right\rangle _{U_{\Lambda}^{0}}\left(
\beta ,\mu
_{\Lambda }^{0}\left(\beta, \rho \right) \right)  \nonumber \\
&&\times \stackunder{k\in \mathcal{K}_{2d}}{\prod }
\left\langle e^{\frac{i}{\sqrt{2V}}\left( \overline{h_{k}}
a_{k}+h_{k}a_{k}^{*}\right) }\right\rangle _{T_{\Lambda}}\left(
\beta ,\mu
_{\Lambda }^{0}\left(\beta, \rho \right) \right)  \nonumber \\
&&\times \stackunder{k\in \Lambda^{*}\setminus\{\mathcal{K}_{2d}\cup\{0\}\}}
{\prod }
\left\langle e^{\frac{i}{\sqrt{2V}}\left( \overline{h_{k}}%
a_{k}+h_{k}a_{k}^{*}\right) }\right\rangle _{T_{\Lambda}}\left(
\beta ,\mu _{\Lambda }^{0}\left(\beta, \rho \right) \right)
\end{eqnarray}
the limit :
\begin{eqnarray}\label{trunc8}
\widetilde{\mathbb{E}}^{0}\left( \beta ,\rho ;h\right) &\equiv
&\text{ } \stackunder{\Lambda }{\lim }\mathbb{E}_{\Lambda
}^{0}\left( \beta ,\mu
_{\Lambda }^{0}\left(\beta, \rho \right) ;h\right)  \nonumber \\
&=&J_{0}\left( \sqrt{2\rho _{0}^{0}\left( \beta ,0\right) }\left|
\widehat{h}_{0}\right| \right)
\exp \left\{ -\frac{1}{4}\left\| h\right\| ^{2}-\frac{1
}{2}A_{\beta ,0}\left( h,h\right) \right\}  \nonumber \\
&&\times \exp \left\{ -\frac{1}{2}\left| \widehat{h}_{0}\right| ^{2}\left(
\rho -\rho _{c}^{0}\left( \beta \right) \right) \right\},
\end{eqnarray}
which \textit{coincides} with the generating functional (\ref{CCRnew10}),
or (\ref{cond}).
\begin{theorem}
\label{truncTh}
The generating functionals for the models $H_{\Lambda}^{0}$ and $H_{\Lambda}^{I}$
are identical :
\begin{equation}\label{trunc9}
\widetilde{\mathbb{E}}^{0}\left( \beta ,\rho ;h\right) =
\widetilde{\mathbb{E}}^{I}\left( \beta ,\rho ;h\right),
\end{equation}
for any $\beta$ and $\rho$.
\end{theorem}
\textbf{5.2} Notice that the relation between grand-canonical
and canonical generating functionals (\ref{Kac1}) for the
PBG can be seen by
means of the identity :
\begin{equation}\label{identity}
\exp(-\frac{1}{2}\lambda z^2)= \stackunder{0}{\stackrel{ +\infty
}{\int}}\frac{dt}{\lambda} e^{-t/\lambda} J_{0}\left( \sqrt{2t
}\left|z\right| \right),
\end{equation}
for $\lambda>0$.
It yields the representation of the grand-canonical
generating functionals
(\ref{cond}) (or (\ref{trunc8})) via the \textit{Kac measure}:
\begin{equation}\label{hypoth1}
\widetilde{\mathbb{E}}^{I}\left( \beta ,\rho ;h\right)=
\stackunder{\mathbb{R}^1}{\int } K_{\rho,\rho _{c}^{I}\left( \beta \right)}(dx)
\,\mathbb{E}^{I}\left( \beta ,\mu=0; h\right)
J_{0}\left( \sqrt{2(x-\rho _{c}^{I}\left( \beta \right))}\left|
\widehat{h}_{0}\right| \right).
\end{equation}
Following (\ref{Kac1}) and (\ref{Kac2}) this gives a temptation to identify
\begin{equation}\label{hypoth2}
\mathbb{E}^{I}\left( \beta ,\mu=0; h\right)
J_{0}\left( \sqrt{2(\rho-\rho _{c}^{I}\left( \beta \right))}\left|
\widehat{h}_{0}\right| \right)=
\mathbb{E}_{can}^{I}\left( \beta ,\rho ;h\right),
\end{equation}
for $\rho>\rho _{c}^{I}\left( \beta \right)$, with the generating functional
$\mathbb{E}_{can}^{I}\left( \beta ,\rho ;h\right)$ of
the model $H_{\Lambda}^{I}$ (or $H_{\Lambda}^{0}$) in the \textit{canonical}
ensemble. But actually we do not know neither
$\mathbb{E}_{can}^{I}\left( \beta ,\rho ;h\right)$, nor
$\mathbb{E}_{can}^{0}\left( \beta ,\rho ;h\right)$, for
$\rho>\rho _{c}^{I}\left( \beta \right)$, to check this hypothesis.

For $\rho\leq\rho _{c}^{I}\left( \beta \right)$ the representation
of the grand-canonical generating functional via the degenerate
\textit{Kac measure} (\ref{Kac0}) and the canonical generating functional
follows directly from (\ref{CCR45}), or (\ref{trunc2}).
\newline
\textbf{5.3}  In the present paper we are not concerned with the explicit
form of the CCR representations corresponding to the generating
functional $\widetilde{\mathbb{E}}^{I}\left( \beta ,\rho ;h\right)$
for different densities $\rho$. Therefore, we limit
ourselves only by few remarks : \\
- In domain $\rho\leq\rho(\beta,\varepsilon_0)$ one has:
\begin{equation}\label{genfunc1}
\widetilde{\mathbb{E}}^{I}\left( \beta ,\rho <\rho _{c}^{I}\left(
\beta \right) ;h\right) =\exp \left\{ - \frac{1}{4}\left\|
h\right\| ^{2}-\frac{1}{2}A_{\beta ,\mu ^{I}\left(\beta, \rho
\right)<0 }\left( h,h\right) \right\},
\end{equation}
where quadratic form $\frac{1}{4}\left\| h\right\|
^{2}+\frac{1}{2}A_{\beta ,\mu ^{I}\left(\beta, \rho \right)
}\left( h,h\right)$ is \textit{closable}. It is known that in this
case the representation is a \textit{factor} corresponding to the
class of \textit{quasi-free} states, see \cite{ManuceauVerbeure}
and \cite{BrattelliRobinson} for
details.  \\
- By virtue of (\ref{identity}) the
$\widetilde{\mathbb{E}}^{I}\left( \beta ,\rho ;h\right)$ in domain
$\rho(\beta,\varepsilon_0)<\rho < \rho _{c}^{I}\left( \beta
\right)$ is the (inverse) Laplace transform of a generating
functional with the \textit{non-closable} quadratic form
$\frac{1}{2}\lambda \left| \widehat{h}_{0}\right|^2 +
\frac{1}{4}\left\| h\right\| ^{2}+\frac{1}{2}A_{\beta ,\mu
^{I}\left(\beta, \rho \right) }\left( h,h\right)$. Hence, the
representation is \textit{not} a
factor.\\
- Because of the term $-\frac{1}{2}\left| \widehat{h}_{0}\right|
^{2}\widetilde{\rho }_{0}^{I}\left( \beta ,\rho \right)$, see
(\ref{cond}), the same remark is valid for representation in
domain $\rho\geq\rho _{c}^{I}\left( \beta \right)$. For details of
construction of the representation in these cases see
\cite{LewisPule1}.

\appendix

\let\oldsect\section
\def\section#1{\def\thesection{Appendix \Alph{section}}
\oldsect{#1}\def\thesection{\Alph{section}}}

\setcounter{proposition}{0}
%

\Section{\label{AppendixCCR A}}

The aim of this appendix is to investigate the \textit{type III}
BE\ condensate in the model (\ref{diagmodel1}), and to add some
new results to what is known since
\cite{BruZagrebnov5,BruZagrebnov7}. The essential point is to
obtain the asymptotics of $\mu _{\Lambda }^{I}\left( \rho \right)
$, which is the solution the of equation
\begin{equation}
\rho=\rho _{\Lambda }^{I}\left( \beta ,\mu _{\Lambda }^{I}\left(\beta, \rho
\right) \right) =\frac{1}{V}\stackunder{k\in \Lambda
^{*}\backslash \left\{ 0\right\} }{\sum }\left\langle
N_{k}\right\rangle _{H_{\Lambda }^{I}}\left(
\beta ,\mu _{\Lambda }^{I}\left(\beta, \rho \right) \right) +\left\langle \frac{%
N_{0}}{V}\right\rangle _{H_{\Lambda }^{I}}\left( \beta ,\mu
_{\Lambda }^{I}\left(\beta, \rho \right) \right) ,
\label{CCR30bis}
\end{equation}
see (\ref{Generating4.0}). We recall that by (\ref{Generating0}) one has
\[
\stackunder{\Lambda }{\lim }\left\langle
\frac{N_{0}}{V}\right\rangle _{H_{\Lambda }^{I}}\left( \beta ,\mu
_{\Lambda }^{I}\left(\beta, \rho \right) \right) =\rho _{0}^{I}\left(
\theta ,\mu ^{I}\left( \beta,\rho \right) \right) ,
\]
where we put $\mu ^{I}\left(\beta, \rho \right) \equiv
\stackunder{\Lambda }{\lim }\mu _{\Lambda }^{I}\left(\beta, \rho
\right) $. To simplify the arguments, we consider below a
\textit{cubic} box $\Lambda \subset \mathbb{R}^{d>2}$ of the volume
$V=\left| \Lambda \right| =L^{d}$, and (at the very last moment)
we put all constants $\{g_k\}_{ k\in\Lambda^{*}}$ to be equal to
$g >0$ .
\newline
$\mathbf{A.1}$ Let $\widetilde{D}_{-}^{\left( \Lambda
\right) }$ and $\widetilde{D}_{+}^{\left( \Lambda \right) }$ be two sets
defined by
\begin{equation}
\begin{array}{l}
\widetilde{D}_{-}^{\left( \Lambda \right) }\equiv \left\{ k\in \Lambda
^{*}\backslash \left\{ 0\right\} :\varepsilon _{k}-\mu _{\Lambda }^{I}\left(\beta,
\rho \right) -\frac{g_{k}}{V}\leq 0\right\} \\
\widetilde{D}_{+}^{\left( \Lambda \right) }\equiv \left\{ k\in \Lambda
^{*}\backslash \left\{ 0\right\} :\varepsilon _{k}-\mu _{\Lambda }^{I}\left(\beta,
\rho \right) -\frac{g_{k}}{V}>0\right\},
\end{array}
\label{CCRnew1bisbis}
\end{equation}
cf. (\ref{CCRnew1bisbisbis}). Then (\ref{CCR30bis}) transforms into
\begin{equation}
\rho=\frac{1}{V}\stackunder{k\in \widetilde{D}_{-}^{\left( \Lambda \right) }}{
\sum }\left\langle N_{k}\right\rangle _{H_{\Lambda }^{I}}
\left( \beta ,\mu _{\Lambda }^{I}\left(\beta,
\rho \right) \right)+\frac{1}{V}
\stackunder{k\in \widetilde{D}_{+}^{\left( \Lambda \right) }}{\sum }
\left\langle N_{k}\right\rangle _{H_{\Lambda }^{I}}\left( \beta ,
\mu _{\Lambda }^{I}\left(\beta,
\rho \right) \right)+\left\langle \frac{N_{0}
}{V}\right\rangle _{H_{\Lambda }^{I}}\left( \beta ,\mu _{\Lambda }^{I}\left(\beta,
\rho \right) \right).  \label{diagmodel74}
\end{equation}
Notice that, since $\varepsilon _{k\neq0}=O(V^{-2/d})$ for $d>2$,
the set $\widetilde{D}_{-}^{\left( \Lambda \right)}=\emptyset$,
if $\mu _{\Lambda }^{I}\left(\beta,
\rho \right)\leq 0$ for large $V$.
\begin{lemma}
\label{diagmodel-lm1}\cite{BruZagrebnov5} Let $g_{+}\geq g_{k}\geq g_{-}>0$
for $k\in \Lambda ^{*}\backslash \left\{ 0\right\} $. Then for any $%
\varepsilon _{0}\in \mathbb{R}^{1}$ and $k\in \widetilde{D}_{+}^{\left( \Lambda
\right) }$, one has the estimate:
\begin{equation}
\left\langle N_{k}\right\rangle _{H_{\Lambda }}\leq \frac{1}{e^{\beta \left(
\varepsilon _{k}-\mu _{\Lambda }^{I}\left(\beta, \rho \right) -\frac{g_{k}}{V}%
\right) }-1},  \label{diagmodel49}
\end{equation}
for $V$ sufficiently large.
\end{lemma}
\textit{\textit{Proof.} }By the correlation inequalities for the Gibbs state
$\omega _{\Lambda }^{I}\left( -\right) \equiv \left\langle -\right\rangle
_{H_{\Lambda }^{I}}$ (see \cite{FannesVerbeure2,FannesVerbeure3}):
\begin{equation}
\beta \omega _{\Lambda }^{I}\left( X^{*}\left[ H_{\Lambda }^{I}\left( \mu
\right) ,X\right] \right) \geq \omega _{\Lambda }^{I}\left( X^{*}X\right) 
\ln
\frac{\omega _{\Lambda }^{I}\left( X^{*}X\right) }{\omega _{\Lambda
}^{I}\left( XX^{*}\right) },  \label{diagmodel50}
\end{equation}
where $X$ is an observable from the domain of the commutator $\left[
H_{\Lambda }^{I}\left( \mu \right) ,.\right] $ with $H_{\Lambda }^{I}\left(
\mu \right) \equiv H_{\Lambda }^{I}-\mu N_{\Lambda }$, we can deduce
\begin{equation}
\beta \omega _{\Lambda }^{I}\left( a_{k}^{*}\left[ H_{\Lambda }^{I}\left(
\mu \right) ,a_{k}\right] \right) \geq \omega _{\Lambda }^{I}\left(
N_{k}\right) \ln \frac{\omega _{\Lambda }^{I}\left( N_{k}\right) }{\omega
_{\Lambda }^{I}\left( N_{k}\right) +1},  \label{diagmodel51}
\end{equation}
for $X=a_{k}$ ($k\neq 0$). Since for $k\neq 0$,
\[
\left[ H_{\Lambda }^{I}\left( \mu \right) ,a_{k}\right] =\left( \mu
-\varepsilon _{k}\right) a_{k}-\frac{g_{k}}{V}a_{k}^{*}a_{k}^{2},
\]
from (\ref{diagmodel51}) one finds for $\mu =\mu _{\Lambda }^{I}\left( \rho
\right) $
\[
\beta \omega _{\Lambda }^{I}\left( \left( \mu _{\Lambda }^{I}\left( \rho
\right) +\frac{g_{k}}{V}-\varepsilon _{k}\right) N_{k}-\frac{g_{k}}{V}%
N_{k}^{2}\right) \geq \omega _{\Lambda }^{I}\left( N_{k}\right) \ln \frac{%
\omega _{\Lambda }^{I}\left( N_{k}\right) }{\omega _{\Lambda }^{I}\left(
N_{k}\right) +1},
\]
which gives:
\begin{equation}
\beta \left( \varepsilon _{k}-\mu _{\Lambda }^{I}\left( \rho \right) -\frac{%
g_{k}}{V}\right) \omega _{\Lambda }^{I}\left( N_{k}\right) \leq \omega
_{\Lambda }^{I}\left( N_{k}\right) \ln \frac{\omega _{\Lambda }^{I}\left(
N_{k}\right) +1}{\omega _{\Lambda }^{I}\left( N_{k}\right) }.
\label{diagmodel52}
\end{equation}
Now the rest of the proof is essentially due to solution for $x\geq 0$ of
the inequality (\ref{diagmodel52}):
\begin{equation}
b_{k}\leq \ln \frac{x+1}{x},  \label{diagmodel53}
\end{equation}
where
\begin{equation}
x\equiv \omega _{\Lambda }^{I}\left( N_{k}\right) \text{, }b_{k}\equiv \beta
\left( \varepsilon _{k}-\mu _{\Lambda }^{I}\left( \rho \right) -\frac{g_{k}}{%
V}\right) .  \label{diagmodel54}
\end{equation}
Since for $k\in \widetilde{D}_{+}^{\left( \Lambda \right) }$ we have $%
b_{k}>0 $, the inequality (\ref{diagmodel53}) implies (\ref{diagmodel49})
for sufficiently large $\Lambda $. $\blacksquare $
\begin{remark}
\label{remarkApp1}\cite{BruZagrebnov5} Because of repulsive interaction
$g_{k}\geq g_{-}>0$
for $k\in \Lambda ^{*}$, see (\ref{diagmodel2}), the finite-volume pressure
$p _{\Lambda }^{I}\left(\beta,\mu \right)$ of the model (\ref{diagmodel1})
is a convex function of $\mu \in \mathbb{R}^{1}$ such that
$\lim_{\Lambda}p _{\Lambda }^{I}\left(\beta,\mu>0 \right)=+\infty$.
This means that (in contrast to the PBG)
for large $\rho$ the solution $\mu _{\Lambda }^{I}\left( \rho \right) $
of the equation (\ref{CCR30bis}) should be \textit{positive}. By convexity :
\begin{equation}\nonumber
\frac{p _{\Lambda }^{I}\left(\beta,\mu \right)-
p _{\Lambda }^{I}\left(\beta,\mu=0 \right)}{\mu}\leq
\partial_{\mu}p _{\Lambda }^{I}\left(\beta,\mu \right)=
\rho _{\Lambda }^{I}\left(\beta,\mu \right)
\end{equation}
for $\mu>0$, one gets that
$\lim_{\Lambda}\rho _{\Lambda }^{I}\left(\beta,\mu>0 \right)=+\infty$.
Therefore, $\lim_{\Lambda}\mu _{\Lambda }^{I}\left( \rho \right)\leq 0 $,
i.e., the \textit{positive} solution of (\ref{CCR30bis}) must
go to zero in the thermodynamic limit.
\end{remark}
\begin{corollary}
\label{corollaryCor1} For $\rho >\rho_{c}^{I}\left( \beta \right)$ and $V$
sufficiently large the solution $\mu _{\Lambda }^{I}\left(\beta, \rho
>\rho_{c}^{I} \left( \beta \right) \right)> 0 $.
\end{corollary}
\textit{Proof.} Suppose that $\mu _{\Lambda }^{I}\left(\beta,
\rho \right)\leq 0 $ for any $\rho$ and $\Lambda$.
Then $\widetilde{D}_{-}^{\left( \Lambda \right)}=\emptyset$, i.e.,
$\widetilde{D}_{+}^{\left( \Lambda \right) }= \Lambda
^{*}\backslash \left\{ 0\right\}$, and by Lemma \ref{diagmodel-lm1}
one gets the estimate:
\begin{eqnarray}
&&\frac{1}{V}\stackunder{k\in \Lambda
^{*}\backslash \left\{ 0\right\} }{\sum }\left\langle
N_{k}\right\rangle _{H_{\Lambda }^{I}}\left(
\beta ,\mu _{\Lambda }^{I}\left(\beta, \rho \right) \right) +\left\langle \frac{
N_{0}}{V}\right\rangle _{H_{\Lambda }^{I}}\left( \beta ,\mu
_{\Lambda }^{I}\left(\beta, \rho \right) \right) \leq \nonumber \\
&&\frac{1}{V}\stackunder{k\in \Lambda
^{*}\backslash \left\{ 0\right\} }{\sum }\frac{1}{e^{\beta \left(
\varepsilon _{k}-\mu _{\Lambda }^{I}\left(\beta, \rho \right) -\frac{g_{k}}{V}
\right) }-1} +\left\langle \frac{
N_{0}}{V}\right\rangle _{H_{\Lambda }^{I}}\left( \beta ,\mu
_{\Lambda }^{I}\left( \beta,\rho \right) \right).\nonumber
\end{eqnarray}
Then, by virtue of $\mu _{\Lambda }^{I}\left(\beta, \rho \right)\leq 0 $, we get
the following inequalities in the thermodynamic limit:
\begin{eqnarray}
&& \rho = \lim_{\Lambda}\frac{1}{V}\stackunder{k\in \Lambda^{*}}{\sum }\left\langle
N_{k}\right\rangle _{H_{\Lambda }^{I}}\left(
\beta ,\mu _{\Lambda }^{I}\left(\beta, \rho \right) \right)\leq \nonumber \\
&& \lim_{\Lambda}\rho
_{\Lambda }^{P}\left(\beta,\mu
_{\Lambda }^{I}\left(\beta, \rho \right) \right) + \stackunder{\Lambda }
{\lim }\left\langle
\frac{N_{0}}{V}\right\rangle _{H_{\Lambda }^{I}}\left( \beta ,\mu
_{\Lambda }^{I}\left(\beta, \rho \right) \right)\leq  \nonumber \\
&& \rho_{c}^{P}\left(\beta\right)+\rho_{0}^{I}\left(\beta,
\mu=0\right)=\rho_{c}^{I}\left(\beta\right),
\end{eqnarray}
see (\ref{Generating0})-(\ref{Generating1}). But this is impossible for $\rho >\rho
_{c}^{I}\left( \beta \right)$, that proves the assertion. \,\,$\blacksquare $

\noindent Therefore, Remark \ref{remarkApp1} and  Corollary
\ref{corollaryCor1} state that
\begin{equation} \label{limitmu}
\lim_{\Lambda}\mu _{\Lambda }^{I}\left(\beta, \rho >\rho
_{c}^{I}(\beta) \right)=0.
\end{equation}
\begin{corollary}
\label{corollaryCor2} The representation (\ref{diagmodel74}), together with
arguments of Corollary \ref{corollaryCor1} and (\ref{limitmu}), allow to
refine the localisation of the \textit{non-extensive} BE condensation
(\ref{Generating2}):
\begin{eqnarray} \label{nonext1}
&&\widetilde{\rho }_{0}^{I}\left( \beta ,\rho \right) =
\lim_{\Lambda}\frac{1}{V}\stackunder{k\in
\widetilde{D}_{-}^{\left( \Lambda \right) }}{%
\sum }\left\langle N_{k}\right\rangle _{H_{\Lambda }^{I}}
\left( \beta ,\mu _{\Lambda }^{I}\left(\beta,
\rho \right) \right)\nonumber \\
&&= \rho - \lim_{\Lambda}\frac{1}{V}
\stackunder{k\in \widetilde{D}_{+}^{\left( \Lambda \right) }}{\sum }%
\left\langle N_{k}\right\rangle _{H_{\Lambda }^{I}}\left( \beta ,
\mu _{\Lambda }^{I}\left(\beta,
\rho \right) \right)- \lim_{\Lambda}\left\langle \frac{N_{0}
}{V}\right\rangle _{H_{\Lambda }^{I}}\left( \beta ,
\mu _{\Lambda }^{I}\left(\beta,
\rho \right) \right)\nonumber \\
&&=\rho - \rho_{c}^{I}\left(\beta\right).
\end{eqnarray}
\end{corollary}
$\mathbf{A.2}$ Our next step is to calculate the \textit{asymptotics}
of $\mu _{\Lambda }^{I}\left(\beta, \rho >\rho_{c}^{I}(\beta) \right)$
in (\ref{limitmu}). To this end suppose that for
$V\rightarrow+\infty$ it has the form :
\begin{equation} \label{asymptmu}
\mu _{\Lambda }^{I}\left(\beta, \rho >\rho
_{c}^{I}(\beta) \right)= \frac{B}{V^{\gamma}}+ o\left(V^{-\gamma}\right),
\end{equation}
with $\gamma>0$ and $B>0$  that should be defined from equation (\ref{CCR30bis}).
\begin{remark}
\label{remarkApp2}
Suppose that $\gamma >2/d$. Since $\varepsilon _{k\neq0}=O(V^{-2/d})$ ,
then the set $\widetilde{D}_{-}^{\left( \Lambda \right)}=\emptyset$, for
large $V$. Therefore, the same line of reasoning as in Corollary
\ref{corollaryCor1} produces
a contradiction to our main assumption : $\rho >\rho_{c}^{I}(\beta)$.
Hence, we must have :
\begin{equation}\label{gamma}
\gamma\leq 2/d.
\end{equation}
\end{remark}
By virtue of additive structure of the Hamiltonian (\ref{CCR20}), for any
$k\in\Lambda^{*}$ we get that
\begin{equation}\label{mean}
\left\langle \frac{N_{k}}{V}\right\rangle
_{H_{\Lambda}^{I}}\left(\beta, \mu _{\Lambda }^{I}(\beta, \rho
)\right)=\left\langle \frac{N_{k}}{V}\right\rangle
_{H_{k}^{I}}\left(\beta, \mu _{\Lambda }^{I}(\beta, \rho )\right)
=\frac{1}{V}\stackunder{n=0}{ \stackrel{+\infty }{\sum }}\,\, n
\,\, \nu _{\Lambda ,k}\left( \beta,\rho; n \right),
\end{equation}
with probability measures : $\{ \nu _{\Lambda ,k}\left( \beta,\rho\,
;n \right) \}_ {\Lambda\subset\mathbb{R}^{d} ,\, k\in\Lambda^{*}}$:
\begin{equation}\label{measure}
\nu _{\Lambda ,k}\left( \beta,\rho\, ;n \right) \equiv\frac{e^{-\beta \left[
\left(
\varepsilon _{k}-\mu _{\Lambda }^{I}\left(\beta, \rho \right) -\frac{g_{k}}{2V}
\right) n+\frac{g_{k}}{2V}n^{2}\right] }}{\stackunder{n=0}{
\stackrel{+\infty }{\sum }} e^{-\beta \left[ \left( \varepsilon
_{k}-\mu
_{\Lambda }^{I}\left(\beta, \rho \right) -\frac{g_{k}}{2V}\right) n+\frac{g_{k}}{2V}
n^{2}\right]}}.
\end{equation}
{From} (\ref{measure}) it is clear that we have to distinguish two
domains :
\begin{equation}
\begin{array}{l}
D_{-}^{\left( \Lambda \right) }\equiv \left\{ k\in \Lambda
^{*}\backslash
\left\{ 0\right\} :\varepsilon _{k}-\mu _{\Lambda }^{I}\left(\beta, \rho \right) -%
\frac{g_{k}}{2V}< 0\right\} , \\
D_{+}^{\left( \Lambda \right) }\equiv \left\{ k\in \Lambda
^{*}\backslash
\left\{ 0\right\} :\varepsilon _{k}-\mu _{\Lambda }^{I}\left(\beta, \rho \right) -%
\frac{g_{k}}{2V}\geq 0\right\} ,
\end{array}
\label{CCRnew1bis}
\end{equation}
cf. (\ref{CCRnew1bisbisbis}) and (\ref{CCRnew1bisbis}). Since
$D_{-}^{\left( \Lambda \right) }\subset\widetilde{D}_{-}^{\left(
\Lambda \right) }$, our next statement makes the localisation of
the \textit{non-extensive} BE condensation more precise, cf.
(\ref{nonext1}).
\begin{lemma}
\label{theoremCCR1}For $\rho >\rho _{c}^{I}\left( \beta \right) $
one has :
\begin{equation}
\widetilde{\rho }_{0}^{I}\left( \beta ,\rho
\right)=\stackunder{\Lambda }{\lim }\frac{1}{V}\stackunder{k\in
D_{-}^{\left( \Lambda \right) }}{\sum }\left\langle
N_{k}\right\rangle _{H_{\Lambda }^{I}}\left(\beta,\mu _{\Lambda
}^{I}\left(\beta, \rho \right)\right) =\rho -\rho _{c}^{I}\left(
\beta \right) >0.  \label{Generating9}
\end{equation}
\end{lemma}
\noindent \textit{Proof.} By (\ref{nonext1}) it is sufficient to
prove that
\begin{equation}\label{diff}
\stackunder{\Lambda }{\lim }\frac{1}{V}\stackunder{k\in {
\widetilde{D}_{-}^{\left( \Lambda \right)}\backslash D_{-}^{\left(
\Lambda \right)}}}{\sum }\left\langle N_{k}\right\rangle
_{H_{\Lambda }^{I}}\left(\beta,\mu _{\Lambda }^{I}\left(\beta,
\rho \right)\right)= 0 .
\end{equation}
Since $0<g_{-}\leq g_{k}\leq g_{+}$, by (\ref{mean}) and
(\ref{measure}) we get that
\begin{equation}\label{diff1}
\left\langle N_{k}\right\rangle _{H_{\Lambda
}^{I}}\left(\beta,\mu _{\Lambda }^{I}\left(\beta, \rho
\right)\right)  \leq \frac{\stackunder{n=0}{ \stackrel{+\infty
}{\sum }}n e^{-\beta g_{k}n^{2}/2V}}{\stackunder{n=0}{
\stackrel{+\infty }{\sum }}e^{-\beta
g_{k}n^{2}/2V}}\frac{\stackunder{n=0}{ \stackrel{+\infty }{\sum
}}e^{-\beta g_{k}n^{2}/2V}}{\stackunder{n=0}{ \stackrel{+\infty
}{\sum }}e^{-\beta(g_{k}n/2V + g_{k}n^{2}/2V)}} \leq
O\left(V^{1/2}\right)
\end{equation}
for $k\in { \widetilde{D}_{-}^{\left( \Lambda \right)}\backslash
D_{-}^{\left( \Lambda \right)}}$ and large $V$. On the other
hand, the set $
\widetilde{D}_{-}^{\left( \Lambda \right)}\backslash
D_{-}^{\left( \Lambda \right)}$ has the same number of elements as
the set 
\begin{equation}\label{diff2}
\Bigl\{s\equiv\{s_{\alpha}\}_{\alpha=1}^{d}\in\mathbb{Z}^{d}
\backslash\left\{0\right\}\; : \; \frac{g_{-}}{2V} +\mu _{\Lambda
}^{I}\left(\beta, \rho \right)\leq \frac{\hbar
^{2}}{2m}\frac{(2\pi)^{2}}{V^{2/d}}
\stackunder{\alpha=1}{ \stackrel{d}{\sum
}}s_{\alpha}^{2}\leq
\frac{g_{+}}{V} +\mu _{\Lambda }^{I}\left(\beta, \rho
\right)\Bigr\}.
\end{equation}
Since the volume of the elementary cell of the dual lattice
$\Lambda^{*}$ is  equal to $(2\pi)^{d}/V$, the number of points
(\ref{diff2}) for large $V$ is \textit{finite}. Together with the estimate
(\ref{diff1}) this gives (\ref{diff}). \,\,$\blacksquare$
\begin{theorem}
\label{theoremAsymp} Let $g_k=g>0$. If $\rho >
\rho _{c}^{I}\left( \beta \right)$, then asymptotics of
the chemical potential $\mu _{\Lambda }^{I}\left(\beta, \rho
\right)$  has the form (\ref{asymptmu}) with
\begin{equation}\label{B-gamma}
\gamma =2/(d+2)\,\,\, \textrm{and }\,\,\,B(\beta,\rho) = \left( \frac{\rho
-\rho _{c}^{I}\left( \beta \right) }{C}\right) ^{2/(d+2)}.
\end{equation}
Here $C=\left(2m/\hbar ^{2}\right)^{d/2}/\left[g\,\, 2^{d-2}\pi^{d/2}d(d+2)\Gamma(d/2)
\right]>0$, and $\Gamma(z)$ is the Euler gamma function.
\end{theorem}
\noindent \textit{Proof.} One has to tune the values of
$B>0$ and $\gamma>0$ in such a way that to satisfy the equation
(\ref{Generating9}) for $V\rightarrow + \infty$. Since
we have $\gamma\leq 2/d$ (Remark \ref{remarkApp2}), by
using (\ref{mean}) and (\ref{measure}) we get:
\begin{eqnarray} \label{Asymp1}
&&\stackunder{\Lambda }{\lim }\frac{1}{V}\stackunder{k\in
D_{-}^{\left( \Lambda \right) }}{\sum }\left\langle
N_{k}\right\rangle _{H_{\Lambda }^{I}}\left(\beta,\mu _{\Lambda
}^{I}\left(\beta, \rho \right)\right)= \nonumber\\
&&\stackunder{\Lambda }{\lim }\frac{1}{V}\stackunder{\left\{s\in
\mathbb{Z}^{d}\backslash\left\{0\right\} \,:\,  \frac{\hbar
^{2}}{2m}\frac{(2\pi)^{2}}{V^{2/d}}
\stackunder{\alpha=1}{ \stackrel{d}{\sum }}s_{\alpha}^{2}
\leq
{g}/{2V} + B/V^{\gamma}\right\}}{\sum }{\stackunder{n=0}
{ \stackrel{+\infty }{\sum }}}\,\, n \, \nu
_{\Lambda,\, k= {2\pi s}/{V^{1/d}}}\left( \beta,\rho\,; n \right)=
\nonumber\\
&&\stackunder{\Lambda }{\lim }\frac{1}{V}
\stackunder{s\in\mathcal{S_B}}{\sum }\, \stackunder{n=0}
{ \stackrel{+\infty }{\sum }} \,\, n \, \nu
_{\Lambda,\, k= {2\pi s}/{V^{1/d-\gamma/2}}}
\left( V^{1-2\gamma}\beta,\rho\,; \frac{n}{V^{1-\gamma}} \right)=\nonumber\\
&&\stackunder{\Lambda }{\lim }\left\{\frac{V^{1-\gamma}}
{V^{d\gamma/2}}\right\}\frac{1}{V^{1-d\gamma/2}}
\stackunder{s\in\mathcal{S_B}}{\sum }\,\,\stackunder{n=0}
{ \stackrel{+\infty }{\sum }} \frac{n}{V^{1-\gamma}} \, \nu
_{\Lambda,\, k={2\pi s}/{V^{1/d-\gamma/2}}}
\left( V^{1-2\gamma}\beta,\rho\,; \frac{n}{V^{1-\gamma}} \right),
\end{eqnarray}
where $\mathcal{S}_B
\equiv\{s=\left\{s_{\alpha}\right\}_{\alpha=1}^{d}\in
\mathbb{Z}^{d}\backslash\left\{0\right\}: \frac{\hbar
^{2}}{2m}(2\pi)^{2} \stackunder{\alpha=1}{ \stackrel{d}{\sum }}
({s_{\alpha}}/{V^{1/d-\gamma/2}})^{2}\leq B\}$ , see (\ref{S}).

The sum in (\ref{Asymp1}) over $\mathcal{S}_B $  is nothing
but the Darboux-Riemann sum converging to the integral, when
$V\rightarrow+\infty$. Therefore, to get a nontrivial limit in
(\ref{Asymp1}) we must choose the value of $\gamma$ from the condition:
$1-\gamma=d\gamma/2$, i.e., $\gamma=2/(d+2)$ in (\ref{asymptmu}).
For this value of $\gamma$ and $d>2$ one has $1-2\gamma > 0$.
Then the family of the \textit{scaled} probability measures:
\[
\{\nu_{\Lambda ,k={2\pi s}/{V^{1/d-\gamma/2}}}\left(
V^{1-2\gamma}\beta,\rho\, ;n/V^{1-\gamma} \right) \}_
{\Lambda\subset\mathbb{R}^{d} ,\, s\in\mathcal{S_B}},
\]
cf.(\ref{measure}), verifies the Laplace \textit{large deviation
principle} \cite{Lewis1},\cite{LewisPfister1} . Hence, by a \textit{diagonal}
limit involving in (\ref{Asymp1}) the sequence of Darboux-Riemann sums
and probability measures, and by (\ref{Generating9}), we deduce the
equation:
\begin{equation}\label{B}
\rho -\rho _{c}^{I}\left(\beta \right) =\frac{1}{\left(
2\pi \right) ^{d}}\stackunder{\{k:\, \varepsilon_{k} \leq B\}}{\int
}d^{d}k \frac{B-\varepsilon_{k}}{g_k}=
\left\{\left(2m/\hbar ^{2}\right)^{d/2}/\left[g\,\,
2^{d-2}\pi^{d/2}d(d+2)\Gamma(d/2)
\right]\right\}B^{(d+2)/2} ,
\end{equation}
which defines the value of $B=B(\beta,\rho)$.
This finishes the proof of (\ref{B-gamma}). \,\, $\blacksquare$

\let\section\oldsect%

\section*{Acknowledgments}

B.N. thanks the Centre de Physique Th\'{e}orique-Luminy, and its
members for their warm hospitality during the visit that led to
this collaboration. J.-B. B. would like to express his gratitude
to Wolfgang Ludwig Spitzer and Shannon Starr for very useful
comments and remarks. This paper was finished during a stay of V.A.Z. 
at UC Davis. He wishes to thank the Department of Mathematics
of UC Davis for its kind hospitality.
This material is based on work supported by
the National Science Foundation under Grant No. DMS0070774.


\begin{thebibliography}{99}
\bibitem{Einstein}  {\ A. Einstein}, {\ Quantentheorie des einatomigen
idealen Gases}, \newblock {\ Sitzungsberichte der Preussischen
Akademie der Wissenschaften} \newblock I (1925) 3--14.

\bibitem{BruZagrebnov1}  {J.-B. Bru and V.A. Zagrebnov}, {\ Exact phase
diagram of the Bogoliubov Weakly Imperfect Bose gas}, \newblock
{\ Phys. Lett. A} \newblock244 (1998) 371--376.

\bibitem{BruZagrebnov2}  {J.-B. Bru and V.A. Zagrebnov}, {\ Exact solution
of the Bogoliubov Hamiltonian for Weakly Imperfect Bose gas},
\newblock {\ J. Phys. A: Math.Gen.} \newblock31 (1998) 9377--9404.

\bibitem{BruZagrebnov4}  J.-B. Bru and V.A. Zagrebnov, \newblock {\
Thermodynamic Behavior of the Bogoliubov Weakly Imperfect Bose
Gas, in:
Mathematical Results in Statistical Mechanics, eds S. Miracle-Sole and al.} %
\newblock(World Scientific, Singapore, 1999) p. 313-321.

\bibitem{BruZagrebnov6}  {J.-B. Bru and V.A. Zagrebnov}, {\ On condensations
in the Bogoliubov Weakly Imperfect Bose-Gas}, \newblock {\ J. Stat. Phys.} %
\newblock99 (2000) 1297-1338.

\bibitem{Zagrebnov}  {\  V.A. Zagrebnov}, {\ Generalized condensation and
the Bogoliubov theory of superfluidity}, \newblock {\ Condensed
Matter Physics}
\newblock 3 (2000) 265--275.

\bibitem{BruZagrebnov8}  V.A. Zagrebnov and J.-B. Bru, {\ }The Bogoliubov
model of weakly imperfect Bose gas, \newblock {\ }Phys. Rep.
\newblock 350 (5/6) (2001) 291-434.

\bibitem{BruZagrebnov5}  {J.-B. Bru and V.A. Zagrebnov}, {\ Exactly soluble
model with two kinds of Bose-Einstein condensations}, \newblock
{\ Physica A}
\newblock268 (1999) 309--325.

\bibitem{BruZagrebnov7}  {\ J.-B. Bru and V.A. Zagrebnov}, {\ A model with
coexistence of two kinds of Bose condensations}, \newblock {\ J.
Phys. A: Math.Gen.} \newblock33 (2000) 449--464.

\bibitem{ZiffUhlenbeckKac}  {\ R.M. Ziff, G.E. Uhlenbeck and M. Kac}, {\ The
Ideal Bose-Einstein Gas, Revisited}, \newblock {\ Phys. Rep.}
\newblock32(C) (1977) 169--248.

\bibitem{BergLewis2}  {\ M. van den Berg and J.T. Lewis}, {\ On the free
boson gas in a weak external potential}, \newblock {\ Commun. Math. Phys.} %
\newblock81 (1981) 475--494.

\bibitem{BergLewis1}  {\ M. van den Berg and J.T. Lewis}, {\ On generalized
condensation in the free boson gas}, \newblock {\ Physica A}
\newblock110 (1982) 550--564.

\bibitem{Berg1}  {\ M. van den Berg}, {\ On boson condensation into an
infinite number of low-lying levels}, \newblock {\ J. Math. Phys.} \newblock%
23 (1982) 1159--1161.

\bibitem{BergLewisPule}  {\ M. van den Berg, J.T. Lewis, and J.V. Pul\`{e}},
{\ A general theory of Bose-Einstein condensation}, \newblock {\
Helv. Phys. Acta} \newblock59 (1986) 1271--1288.

\bibitem{BergLewisLunn}  {\ M. van den Berg, J.T. Lewis, and M. Lunn},
{\ A general theory of Bose-Einstein condensation and the state
of the free boson gas}, \newblock {\
Helv. Phys. Acta} \newblock59 (1986) 1289--1310.

\bibitem{Pule1}  {\ J.V. Pul\`{e}}, {\ The free boson gas in a weak external
potential}, \newblock {\ J. Math. Phys.} \newblock24 (1983)
138--142.

\bibitem{Berg2}  {\ M. van den Berg}, {\ On condensation in the free-bosons
gas and the spectrum of the Laplacian}, \newblock {\ J. Stat. Phys.} %
\newblock31 (1983) 623--637.

\bibitem{MichoelVerbeure}  {T. Michoel and A. Verbeure}, {\ Non-extensive
Bose-Einstein condensation model}, \newblock {\ J. Math. Phys.}
\newblock40 (1999) 1268--1279.


\bibitem{BergLewisPule1}  {\ M. van den Berg, J.T. Lewis and J.V. Pul\`{e}},
{\ The large deviation principle and some models of an
interacting boson gas}
, \newblock {\ Commun. Math. Phys.} \newblock118
(1988) 61--85.

\bibitem{BergDorlasLewisPule1}  {\ M. van den Berg, T.C. Dorlas, J.T. Lewis
and J.V. Pul\`{e}}, {\ The pressure in the Huang-Yang-Luttinger
model of an interacting boson gas}, \newblock {\ Commun. Math.
Phys.} \newblock128 (1990) 231--245.

\bibitem{DorlasLewisPule2}  T.C. Dorlas, J.T. Lewis and J.V. Pul\`{e}, The
Full Diagonal Model of a Bose Gas, Commun. Math. Phys. 156 (1993)
37--65.

\bibitem{ArakiWoods1}  {H. Araki and E.J. Woods}, {\ Representations of the
Canonical Commutation Relations Describing a Nonrelativistic
Infinite Free Bose Gas}, \newblock {\ J. Math. Phys.} \newblock4
(1963) 637--662.

\bibitem{Cannon1}  {J.T. Cannon}, {\ Infinite Volume Limits of the Canonical
Free Bose Gas States on the Weyl Algebra}, \newblock {\ Commun. Math. Phys.} %
\newblock29 (1973) 89--104.

\bibitem{LewisPule1}  {J.T. Lewis and J.V. Pul\`{e}}, {\ The Equilibrium
States of the Free Boson Gas}, \newblock {\ Commun. Math. Phys.}
\newblock36 (1974) 1--18.

\bibitem{Lewis2}  J.T. Lewis, \newblock The free boson gas in: Mathematics
of Contemporary Physics eds R.F. Streater, p. 209--226. Academic
Press, London and New York, 1972.

\bibitem{HYL}  K. Huang, C.N. Yang, and J.M. Luttinger, \newblock
Imperfect Bose gas with hard-sphere intereactions, Phys. Rev. 105
(1957) 776--784.

\bibitem{BrattelliRobinson}  O.~Brattelli and D.W. Robinson, \newblock {\
Operator Algebras and Quantum Statistical Mechanics, Vol. 2}.
\newblock, 2nd ed : Equilibrium States, models in Quantum
Statistical Mechanics (Springer, New York-London-Paris, 1996).

\bibitem{Lewis1}  J.T. Lewis. \newblock Mark Kac seminar on probability and
physics: The Large Deviation Principle in Statistical Mechanics,
syllabus 17. \newblock Amsterdam, Centrum voor Wiskunde en
Informatica CWI (1985-1987), 1988.

\bibitem{LewisPfister1}  {J.T. Lewis and Ch.-E. Pfister}, {\ Thermodynamic
probability theory: some aspects of large deviations}, \newblock
{\ Russian Math. Surveys} \newblock50 (1995) 279--317.

\bibitem{ManuceauVerbeure}  {J. Manuceau and A. Verbeure}, {\
Quasi-free states of the CCR-algebra and Bogoliubov
transformations}, \newblock {\ Commun. Math. Phys.} %
\newblock9 (1968) 293--302.

\bibitem{FannesVerbeure2}  {M. Fannes and A. Verbeure}, {\ Correlation
Inequalities and Equilibrium States I}, \newblock {\ Commun. Math. Phys.} %
\newblock55 (1977) 125--131.

\bibitem{FannesVerbeure3}  {M. Fannes and A. Verbeure}, {\ Correlation
Inequalities and Equilibrium States II}, \newblock {\ Commun. Math. Phys.} %
\newblock57 (1977) 165--171.
\end{thebibliography}
\end{document}